\documentclass[twocolumn,superscriptaddress,aps,pre]{revtex4-1}

\usepackage{amsmath,amsfonts}
\usepackage{graphicx}

\usepackage{color}

\begin{document}

\title{
Charge migration mechanisms in the DNA at finite temperature revisited;
from quasi-ballistic to subdiffusive transport
}

\author{R.\ P.\ A.\ Lima}
\email{rodrigo@fis.ufal.br}
\affiliation{GISC and GFTC, Instituto de F\'{\i}sica, Universidade Federal de Alagoas, Macei\'{o} AL 57072-970, Brazil}
\author{A.\ V.\ Malyshev}
\email[Corresponding author: ]{a.malyshev@fis.ucm.es}
\affiliation{GISC, Departamento de F\'{\i}sica de Materiales, Universidad Complutense, E-28040 Madrid, Spain}
\affiliation{Ioffe Physical-Technical Institute, St-Petersburg, Russia}

\begin{abstract}

Various charge migration mechanisms in the DNA are studied within the framework of the Peyrard-Bishop-Holstein model which has been widely used to address charge dynamics in this macromolecule. To analyze these mechanisms we consider characteristic size and time scales of the fluctuations of the electronic and vibrational subsystems. It is shown, in particular, that due to substantial differences in these timescales polaron formation is unlikely within a broad range of temperatures. We demonstrate that at low temperatures electronic transport can be quasi-ballistic. For high temperatures, we propose an alternative to polaronic charge migration mechanism: the fluctuation-assisted one, in which the electron dynamics is governed by relatively slow fluctuations of the vibrational subsystem. We argue also that the discussed methods and mechanisms can be relevant for other organic macromolecular systems, such as conjugated polymers and molecular aggregates.

\end{abstract}

\maketitle

\section{Introduction}

Studies of the DNA properties were originally carried out in the field of biology.  However, some important properties of these macromolecules were manifesting signatures of fundamental physical phenomena. In particular, the denaturation of the DNA (complete separation of its strands) was known to have a threshold-like character~\cite{bubbleDynamics}. The latter problem was successfully addressed in 1989 by Peyrard and Bishop (PB) who proposed a model describing the denaturation as a phase transition in this quasi one dimensional (1D) system~\cite{Peyrard1989,Dauxois1993}. However, we would like to remind that the PB model is just a particular case of a more general 1D model of a multi-component nonlinear classical field. The latter was put forward in 1964 by Suris~\cite{Suris64,Suris65} who studied and analyzed its thermodynamic properties in detail, including the specific heat and correlation functions, and predicted that such a model can exhibit a phase transition.  In the particular case of the PB model, the nonlinearity is introduced through the onsite Morse potential which accounts for the elastic energy related to the inter-strand hydrogen bond stretchings. In this case, the phase transition manifests itself in the divergence of these stretchings at the critical temperature, resulting in the denaturation phenomenon. The PB model has been widely used since then because of its success in the description of such an important biological process. 

The interest in the DNA increased considerably after it was advocated as a possible fundamental constituent of a new molecular size electronics (see, for example, reviews \onlinecite{Endres2004,Chakraborty2007} and references therein). Since then the investigation of the DNA became truly interdisciplinary involving extensive studies of its optical\cite{Charra2003,Diaz2007,Sonmezoglu2011,Schimelman2015}, electronic and transport properties\cite{Enrique09,Diaz2008,Ojeda2009,Porath2000,Okahata1998,Fink1999,Rakitin2001,Legrand2006,Yoo2001,Hwang2002,Xu2004,Cohen2005,Braun1998,Storm2001,Cuniberti2002}. Consecutive aromatic rings comprising the DNA double helix are coupled due to the overlap of their $\pi$ orbitals, which determines different important electronic properties of the chain \cite{Murphy1993}.  The original conjecture was that these coupled states could provide a channel for
electric current, converting the DNA in a conducting nanowire, which was very promising for various applications in the field of molecular electronics. 

The DNA electrical properties have been measured experimentally revealing  contradictory results, including
insulating~\cite{Braun1998,Storm2001}, semiconducting~\cite{Cuniberti2002,Yoo2001,Xu2004,Cohen2005} and metallic \cite{Okahata1998,Fink1999,Rakitin2001,Legrand2006} behavior.  Later, it was understood that these differences were probably related to different experimental conditions, such as, sample and contact types, measurement techniques, the DNA environment, etc.  Considerable effort has also been put into theoretical description of the electronic transport in the DNA.  Both purely ballistic transport~\cite{Malyshev2007,Chakraborty2007} and the phonon-related one~\cite{Kalosakas2003a,Kalosakas2011,Maniadis2003,Fuentes2004,Maniadis2005,Starikov2005,Malyshev2009,Enrique09,Gutierrez2006,Wei2008,Chetverikov2016,Astakhova2014} were investigated thoroughly. In the later case there are two opposite limits: those of weak and strong electron-phonon interaction.  In the former case the transport is determined by the electron hopping assisted by phonons, thus, a disordered chain insulating at zero temperature can become conducting at non-zero temperature because of the thermal activation of the hopping mechanism (see, for example, Ref.~\onlinecite{Malyshev2009} and references therein). On the other hand, a strong electron-phonon coupling can lead to the formation of polarons governing the charge migration.~\cite{Maniadis2005,Yu2001,Alexandre2003}. To address the later case, it was proposed that the PB model could be extended to investigate charge migration properties by considering a coupling of the lattice and electronic degrees of freedom~\cite{Dauxois1993,Komineas2002,Maniadis2005}. In this paper we are using such a model.

A polaron, being an electron dressed by virtual phonons, has a larger effective mass than a bare electron~\cite{Landau1948,Landau1965}, which is detrimental for the charge mobility.  Although lower mobility is unfavorable for transport efficiency, electron-phonon interaction should be taken into account, provided it is relevant for the studied system properties.  At zero temperature, a polaronic solution constitutes the ground state of the system.  However, the contribution of the polaronic configuration to the partition function decreases for finite temperature, and it has been an essentially unexplored question, whether the set of configurations close to such a state is statistically meaningful. In this paper, we address this question, studying the polaron dynamics and its lifetime within a broad range of temperatures. We also discuss and analyze an alternative charge migration mechanisms: the fluctuation-assisted one.

The paper is organized as follows. In the next two sections, we revisit the Peyrard-Bishop-Holstein (PBH) model which addresses a coupled electronic and DNA ``lattice" dynamics and calculate its minimum energy configurations at $T=0$. In the following sections, we consider the system dynamics at non zero temperature, both below and above the environment freezing point, for the following very different initial conditions. In Sec.~\ref{sectionPolaron}, we use traditional and somewhat artificial initial condition under which the system in its polaronic state at zero temperature is subjected to the action of a heat bath, {\it i. e.}, we address polaron dynamics during the chain thermalization (heating up) and analyze polaron lifetime. In Sec.~\ref{sectionElectron}, a more realistic initial condition is considered when an electron is injected in a thermalized DNA chain and the coupled system dynamics is studied. In the later case, we discuss different new non-polaronic charge migration regimes, in particular, quasi-ballistic and sub-diffusive ones. Conclusions summarize the paper. 

\section{Model and formalism.}

In this section, we present the minimal set of equations of motion describing the dynamics of an electron-phonon coupled system in the framework of the PBH model (for more details see Refs.~\onlinecite{Kalosakas2003a,Komineas2002,Dauxois1993}). The dynamics of the system can be obtained within the semiclassical approximation: the electronic part is treated quantum-mechanically, while the lattice dynamics is described classically within the framework of the Langevin approach. The following equation of motion of the $n$-th hydrogen bond stretching $y_n$ (which describes the relative motion of bases in a base pair) is used:
\begin{eqnarray}
\nonumber
\mu \frac{d^2y_n}{dt^2}= -\frac{d\,V(y_n)}{d y_n}-\frac{d\,W(y_n,y_{n-1})}{d y_n}
-\frac{d\,W(y_{n+1},y_{n})}{d y_n}
\\
-\mu\gamma\frac{d\,y_n}{dt}-\chi\,|\psi_n|^2+f_n(t)\ ,
\qquad
\label{eq:Hlat}
\end{eqnarray}
where
\begin{align}
 V(y)&= V_0 \left(1-e^{-\alpha y}\right)^2
 \label{Morse}
\end{align}
is the onsite Morse potential and
\begin{align}
 W(y,y^\prime)&= \frac{K\,(y-y^\prime)^2}{2}\left[1 + 
\rho\, \mathrm{e}^{-\beta(y+y^\prime)}\right]
\label{W}
\end{align}
is the nearest-neighbor interaction potential which couples the dynamics of consecutive DNA base pairs.  In Eqs. (\ref{eq:Hlat}-\ref{W}), $\mu$ is the reduced mass of a base pairs (as in the bulk of the literature we consider all reduced masses to be the same), $\gamma$ is the friction constant, $\chi$  is the on-site electron-phonon coupling strength, while $V_0$, $\alpha$, $K$, $\rho$ and $\beta$ are parameters characterizing the elastic potential energy.  These parameters can be obtained through the comparison to the first-principle calculations~\cite{Peyrard1989,Dauxois1993,Dauxois1993b} and experimental data~\cite{Gao1984,Campa1998}. 

Finally, the force $f_n(t)$ describes random action of the environment; it has the following
statistical properties:
\begin{eqnarray}
\nonumber
\langle f_n(t)\rangle= 0\ ,
\\
\langle f_n(t)f_{n^\prime}(t')\rangle=2\gamma\,\mu\, k_B T \delta_{n\,n^\prime}\delta(t-t')\ ,
\label{rand}
\end{eqnarray}
where $k_B$ is the Boltzmann constant and $T$ is the temperature. Such an approach is a microscopic model characterizing a water-type environment in terms of the collisions between the system and the ``fast" reservoir molecules~\cite{Gardiner1986,Kalosakas2003a}. 
%

The original PB model~\cite{Peyrard1989} of the DNA ``lattice" dynamics corresponds to Eq. (\ref{eq:Hlat}) without the term ($\chi\,|\psi_n|^2$) which accounts for the coupling to the electronic
subsystem. The dynamics of the latter (within the framework of the adopted Holstein approximation) can be described by the the following Schr\"{o}dinger equation:
\begin{eqnarray}
i\,\hbar\frac{\partial\,\psi_n}{\partial\,t}=
 -J\,(\psi_{n-1}+\psi_{n+1})+\chi\,y_n\,\psi_n \ ,
 \label{eq:Hel}
\end{eqnarray}
where $\psi_n$ is the charge carrier wave function at the $n$-th site and $J$ is
the nearest neighbor electron coupling.  The last term of Eq.~(\ref{eq:Hel})
accounts for the interaction with the DNA ``lattice" where $\chi$ is the 
electron-phonon coupling constant.  

From the point of view of the relatively fast electronic subsystem, the interaction term
\begin{eqnarray}
U_{n}=\chi\, y_n
\label{eq:Hinter}
\end{eqnarray}
in the Hamiltonian of Eq.~(\ref{eq:Hel}) is a random slowly varying onsite potential energy of the charge.  On the other hand, relatively slow base pairs experience the electron-phonon coupling as an action of an external force which is proportional to the local electronic density (see the term $\chi\,|\psi_n|^2$ in Eq.~(\ref{eq:Hlat})).  

Some care should be taken while treating the interaction terms in 
Eqs.~(\ref{eq:Hlat}) and (\ref{eq:Hel}) because they correspond to the 
first order expansion of the electronic energy
with respect to the deformation of the $\pi$ orbital due to the
hydrogen bond elongation $y_n$. 
For large deviations from the equilibrium
position such an expansion can fail.  However, the Morse
potential makes large negative values of $y_n$ very unlikely, while large
positive values of $y_n$, corresponding to broken bonds, give rise to large potential
barriers for the electron.  Broken bonds are known to suppress the
electronic transport just like high potential barriers do, resulting in a
qualitatively correct impact of large positive displacements on the
transport.

Computing the system dynamics requires specialized approach because of the presence of the stochastic term in the lattice equation of motion. In such a case, generic numerical calculation methods can become inaccurate while specialized algorithms of numerical integration of the stochastic differential equations~\cite{nisde,nisde2} tend to give much more reliable results. Therefore, we used the {\bf 3o4s2g} algorithm: the third order Runge-Kutta method with four sub-steps and two random Gaussian number generated in each integration step.

\begin{figure}[t!]
\includegraphics[width=\linewidth]{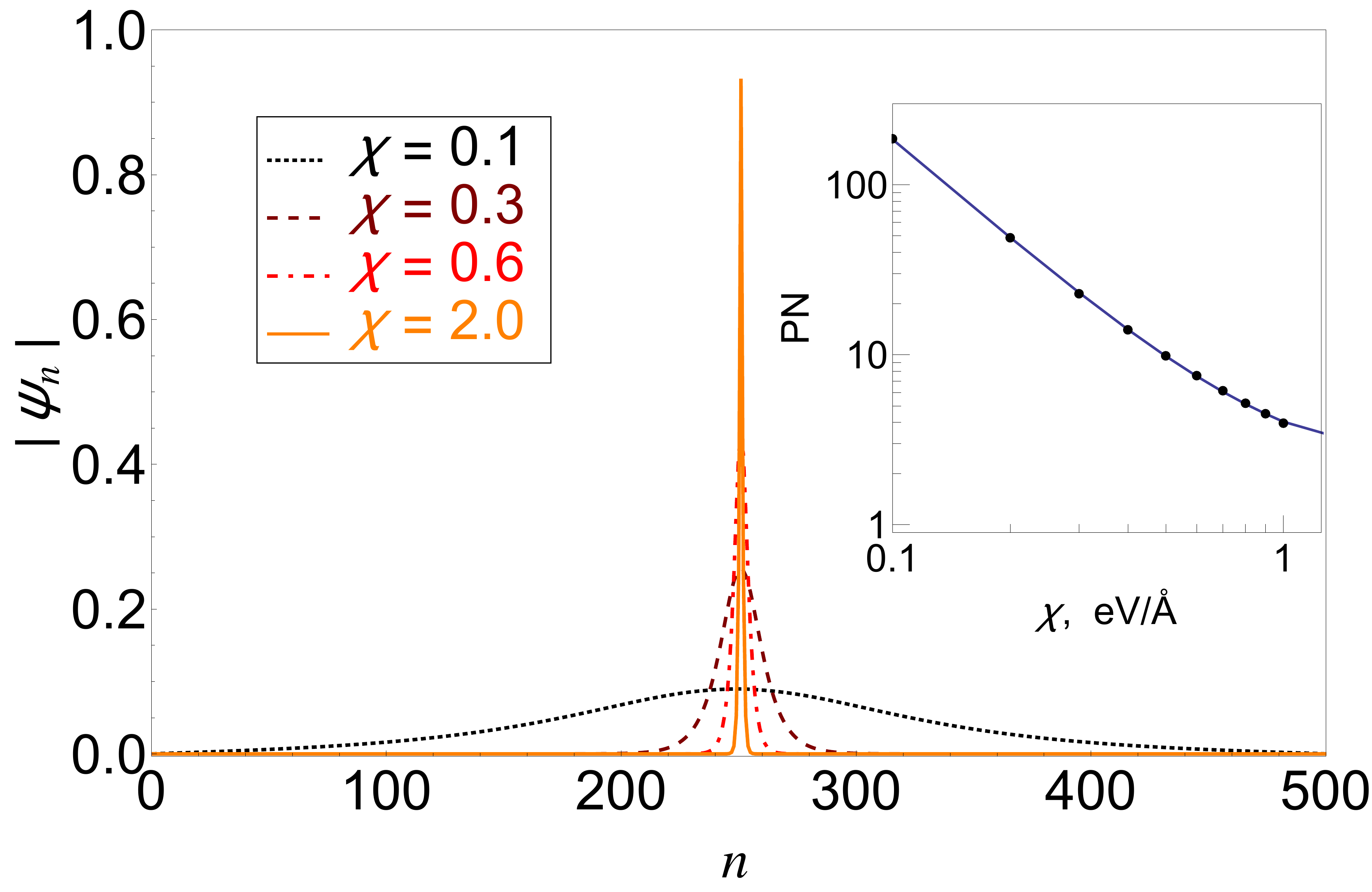}
\caption{
Electronic wavefunction calculated for different values of the coupling constant $\chi$ at $T=0$.  In the inset, the log-log plot of the participation number ($PN$) (see text for the description) versus the coupling $\chi$ is shown; dots give numerical result while the solid line - the best fit function $2.04+1.99\,\chi^{-1.96}$.
}
\label{fig:polaronT0}
\end{figure}

\section{Polaronic solution at $T=0$}

Hereafter we use the following set of parameters (unless stated otherwise): $\mu=300\,\mbox{amu}$, $V_0=0.04\,\mbox{eV}$, $\alpha=4.45\,\mbox{\AA}^{-1}$, $K=0.04\,\mbox{eV/\AA}^{2}$, $\rho=0.5$, $\beta=0.35\,\mbox{\AA}^{-1}$, $\gamma=0.005\,\mbox{ps}^{-1}$, $\chi=0.6\,\mathrm{eV}/\mbox{\AA}$, $J=0.1\,\mbox{eV}$, which are typical values used in literature~\cite{Dauxois1993,Komineas2002,Maniadis2005}.

The minimum energy configuration at zero temperature can be obtained through the iterative process~\cite{Maniadis2003,Diaz2008} or by integrating Eq.~(\ref{eq:Hlat}) with an initial wavefunction $\psi_n$ and displacements $y_n$ which are similar in shape to the expected minimum.  Then, the drag term in Eq.~(\ref{eq:Hlat}) decreases the total energy of the system which finally relaxes to its ground state.  Figure~\ref{fig:polaronT0} shows the electronic wavefunction corresponding to the polaronic ground state for different values of the coupling $\chi$.  

In order to analyze the spacial extent of the electronic state, we use the participation number (PN) defined as:
\begin{equation}
 \mathrm{P}(t) = \left( \sum_{n=1}^N \psi_n^4(t) \right)^{-1}\ .
\label{eq:PN}
\end{equation}
It's physical interpretation is as follows. If the wavefunction is localized at $N^\star$ sites (let $N^\star=\mathrm{const}<N$) then its typical value at a site within it's localization volume is on the order of $1/\sqrt{N^\star}$ and the PN is on the order of $N^\star$. So, it will remain finite in the thermodynamic limit $N\to\infty$. On the other hand, if the wavefunction is extended over the whole system the PN is on the order of the system size $N$ diverging in the thermodynamic limit. Thus, the PN characterizes the spacial extent of the corresponding state and it is a very convenient quantity to study localization/delocalization properties and phenomena. Speaking more strictly, the PN gives the number of sites which provide the main contribution into the normalization of the wavefunction, {\it i. e.} the number of sites where the wavefunction has appreciable amplitude. The later is very important for the electron-phonon interaction too because the coupling term in the lattice equation of motion (\ref{eq:Hlat}) is proportional to the local electronic density $|\psi_n|^2$, so the coupling is significant only at the sites where the wavefunction has appreciable amplitude. In particular, significant negative displacements $y_n$ characteristic for a polaronic configuration can exist only at such sites and therefore the PN can characterize also the size of a polaron if the latter exists. 
If the extent of the wavefunction increases the electron--phonon coupling is reduced,~\cite{Malyshev1998} which can manifest itself in the break-up of a polaron as we show below.

The PN corresponding to the polaronic ground state as a function of the coupling $\chi$ at $T=0$ is shown in the inset of Fig.~\ref{fig:polaronT0}. As expected, the size of the polaron reduces as the interaction grows: the dependence is roughly power-law with the exponent of about $-1.96$ which agrees very well well with a simple analytical estimate giving the exponent $2$. The latter result can be obtained by the variational calculation, using Gaussians for both the displacement $y_n$ and the wavefunction $\psi_n$ in the continuous limit:
\begin{eqnarray}
\nonumber
y(x) &=& y(0)\,\exp{\left(-\frac{x^2}{\lambda^2}\right)}\ , \quad
\nonumber\\
\psi(x) &=& \frac{1}{\pi^{1/4}\lambda^{1/2}}\, \exp{\left(-\frac{x^2}{2\lambda^2}\right)}
\end{eqnarray}
in which case the displacement amplitude $y(0)$, the polaron total energy $E_{p}$, its ``size" $\lambda$ are as follows:
\begin{eqnarray}
\nonumber
y(0) = -\frac{\chi^3}{8\sqrt{2}\pi\,J\,V_0^2\,\alpha^4} \ , \qquad
\nonumber
E_{p} = -\frac{\chi^4}{64\pi\,J\,V_0^2\,\alpha^4} \ , \\
\nonumber
\lambda = \frac{4\sqrt{2\pi}J\,V_0\,\alpha^2}{\chi^2}\  ,
\end{eqnarray}
and the corresponding wavefunction participation number $PN$ is
\begin{eqnarray}
PN = \sqrt{2\pi}\lambda = \frac{8\pi\,J\,V_0\,\alpha^2}{\chi^2}\ .
\end{eqnarray}
The analytical estimate of the $PN$ for our set of parameters is $PN\approx 1.99\,\chi^{-2}$ which is in very good agreement with the numerical result $2.04+1.99\,\chi^{-1.96}$. The discrepancy at large values of $\chi$ is expectable: as the interaction increases and the polaron shrinks in size the continuous approximation becomes inapplicable.

\section{Polaron dynamics during the chain thermalization
\label{sectionPolaron}}

In this section, we study the polaron dynamics during the chain thermalization process using the polaronic solution at zero temperature (as obtained in the previous section) as the initial condition.  We assume also that at $t > 0$ the system is subjected to the action of the heat bath with the temperature $T_0$.  Although the physical relevance of this initial condition could be questioned, it has been widely used in the literature~\cite{Diaz2008,Vidmar2011,Voulgarakis2017,Kalosakas2003a} (and reference therein) and it guarantees the existence of a polaron in the system at the initial moment of time.  If the polaronic solution is relevant (stable) one would expect it to survive the complete thermalization process, until the system reaches the temperature of the bath.  Our calculations show that the polaron falls apart during the thermalization; to account for that we introduce the polaron lifetime, defining it as the polaron break-up or delocalization time $\tau_d$. We study this quantity as a function of the bath temperature $T_0$ and address also the temperature of the DNA chain at $t=\tau_d$, {\it i. e.} at the moment of the polaron break-up.

\begin{figure}
\includegraphics[width=\linewidth]{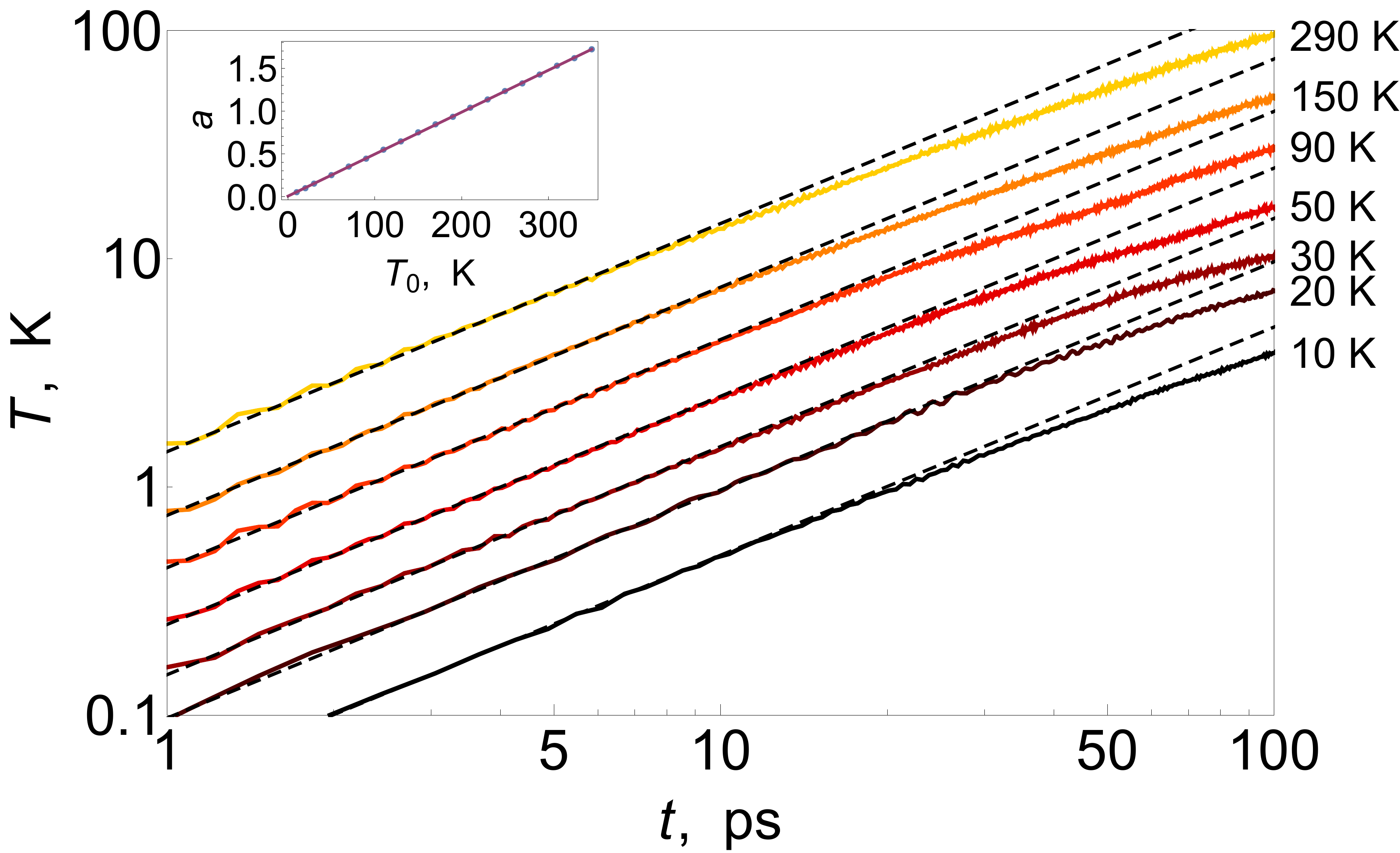}
\caption{
Realization-averaged temperature $T$ of the chain which was initially at rest ($y_n(0)=\dot{y}_n=0$) for different bath temperatures $T_0$. As can be seen from Eq.~(\ref{eq:thermprocess0}), the heating processes is initially linear.  The linear fit to the data (dashed black lines) allow us to obtain the slope $a=T_0/\tau_0$ (whose dependence on the bath temperature is shown in the inset) and therefore, the characteristic time scale $\tau_0 \approx 204$ ps.
}
\label{fig:tauDetermination}
\end{figure}

We found that the thermalization of the chain is governed by a simple exponential law for the current chain temperature $T$ as a function of time:
\begin{equation}
T(t)=T_0\left(1-e^{-t/\tau_0}\right)\ ,
\label{eq:thermprocess}
\end{equation}
where $\tau_0$ gives the time scale of the heating process. The above dependence can be verified numerically. Figure~\ref{fig:tauDetermination} shows time dependencies of the configuration-averaged chain temperature $T$ for different bath temperatures $T_0$.  The temperature of the chain is estimated using the virial theorem, which relates the kinetic energy of the system with its temperature, $E_\mathrm{kin}=N\cdot k_B T/2$ ($k_B$ being the Boltzmann constant).  By fitting the formula
\begin{equation}
T(t) \approx \frac t \tau_0
T_0,\quad t\ll \tau_0, 
\label{eq:thermprocess0}
\end{equation}
to the initial linear part of the curves in Fig.~\ref{fig:tauDetermination}, one obtains the slope $a=T_0/\tau_0$ and therefore the characteristic heating time $\tau_0\approx 204$ ps.  The inset of the figure shows the expected linear dependence of the slope on the bath temperature.

The parameter $\tau_0$ can be estimated analytically by considering the energy transferred to the forced oscillator system. During the initial phase of heating the damping is negligible ($t\ll 1/\gamma$) and the typical displacement is small, so that the interaction potential (\ref{W}) can be neglected too, then a harmonic oscillator model of the Morse potential (\ref{Morse}) is applicable. In such a simplified case the transferred energy at the time $t$ to the forced oscillator is given by~\cite{Landau1969}
\begin{equation}
E_{T}=\frac{1}{2M}
\left|
\int_{-\infty}^t 
\sum_n^N f_n(t^\prime)\, e^{i\,\omega\,t^\prime}\,dt^\prime
\right|^2\ ,
\end{equation}
where $M=N\,\mu$ is the total reduced mass and $\omega$ is the frequency of the harmonic oscillator. Using the statistical properties of the random force $f(t)$, given by Eq.~(\ref{rand}), one can calculate the transferred energy as
$$
E_T=N k_B T\,\gamma\, t\ .
$$
On the other hand, the energy of an oscillator system can be obtained as twice the kinetic energy: $2E_{kin}=N k_B T(t)=N k_B T_0\, t/\tau_0$. Equating the two results gives the estimate of the parameter $\tau_0$ as: 
$$\tau_0=\frac{1}{\gamma}=200\,\mathrm{ps}
$$
which is in a very good agreement with the result obtained numerically from the dependencies in the Fig.~\ref{fig:tauDetermination}: $\tau_0\approx 204$ ps.

\begin{figure}
\includegraphics[width=\linewidth]{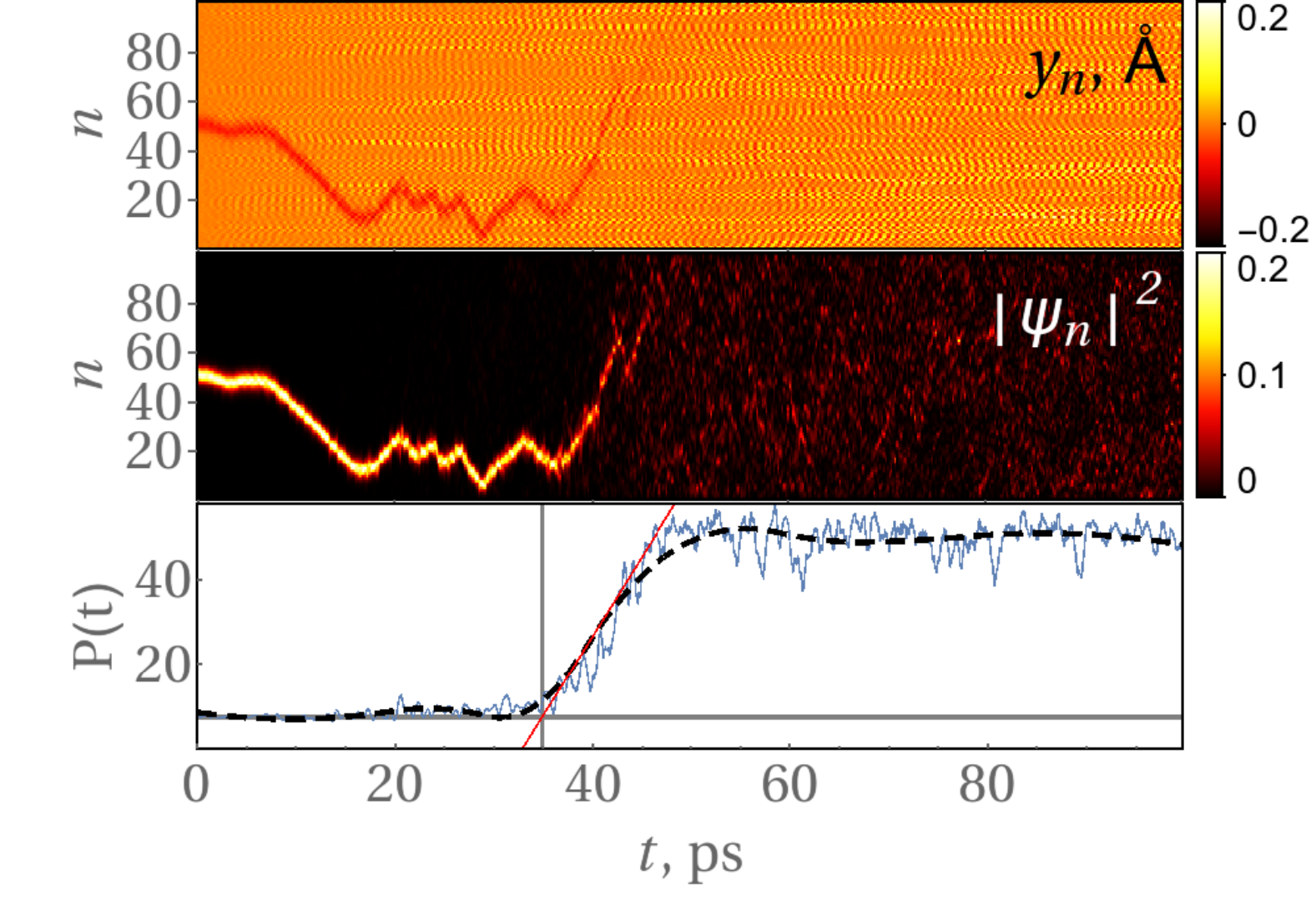}
\caption{
Density plots of the time evolution of the hydrogen bond stretchings $y_n$ (upper panel) and the probability density $|\psi_n|^2$ (middle panel), where $n$ is the number of a base in the DNA chain. The polaron is initially static and located at the center of the chain, at $t=0$ the system is coupled to a heat bath with the temperature $T_0=80$ K.  Lower panel shows the dynamics of the participation number PN. The dashed black and solid red lines in the lower plot are used to calculate the polaron break up time marked with a grey vertical line (see text for details).
}
\label{fig:polaronEvolutionT80}
\end{figure}

Figure~\ref{fig:polaronEvolutionT80} shows an example of the system dynamics for $T_0=80$ K.  The upper plot displays the evolution of bond stretchings $y_n$, the middle one -- the probability density $|\psi_n|^2$ while the lower panel demonstrates the PN for this particular trajectory.  The system starts from a polaronic state which is characterized by highly localized negative stretchings and the wavefunction.  The trace of the polaronic state is clearly visible in the range $t < 40$ ps. Note the clear correspondence of high electronic density and lattice distortion on that part of the trajectory -- a clear fingerprints of a polaron.  As the system is heated up, stronger fluctuations can be observed; their dynamics is governed by small oscillations around the equilibrium position of the Morse potential.  At $t\approx 40$ ps the polaron finally breaks up: the wavefunction delocalizes and spreads over the whole chain while the stretchings become almost random. The lower plot clearly shows the switching between two different regimes, the initial polaronic solution (PN$\simeq 7$) and the final delocalized state (PN$\simeq N/2=50$).

In order to estimate the polaron lifetime or the delocalization time $\tau_d$ -- the time at which the polaron breaks up -- we use several auxiliary quantities proceeding as follows.  First, we calculate the integrated participation number:

\begin{equation}
{\mathrm{I}}(t) = \int_{0}^t \mathrm{P}(t^\prime)\, d t^\prime,
\end{equation}

which is interpolated using a sparse mesh of points giving a smooth function of time $I_{int}(t)$, whose derivative ${\cal{P}}(t)=\dot{I}_{int}(t) $ is a smoothed version of the function ${\mathrm{P}}(t)$. Green dashed line in the lower panel of Fig.~\ref{fig:polaronEvolutionT80} shows the function ${\cal{P}}(t)$.  We define the polaron delocalization $\tau_d$ by the following equation

$$
\dot{{\cal{P}}}(t_{i})=\frac{{\cal{P}}(t_{i})-{\cal{P}}(0)}{t_{i}-\tau_d}\ ,
$$

here $t_{i}$ is the inflection point of ${\cal{P}}(t)$ in the transition region, where the participation number starts growing fast, that is the point where the first derivative $\dot{{\cal{P}}}(t)$ is maximum. Thus, $\tau_d$ is determined by the intersection of the linear interpolation of the ${\cal{P}}(t)$ in the vicinity of the inflection point (see the solid red line in the lower panel of Fig.~\ref{fig:polaronEvolutionT80}) and the line, giving the initial polaron size ${\cal{P}}(0)$ (shown by the horizontal gray line).

\begin{figure}[t]
\includegraphics[width=\linewidth]{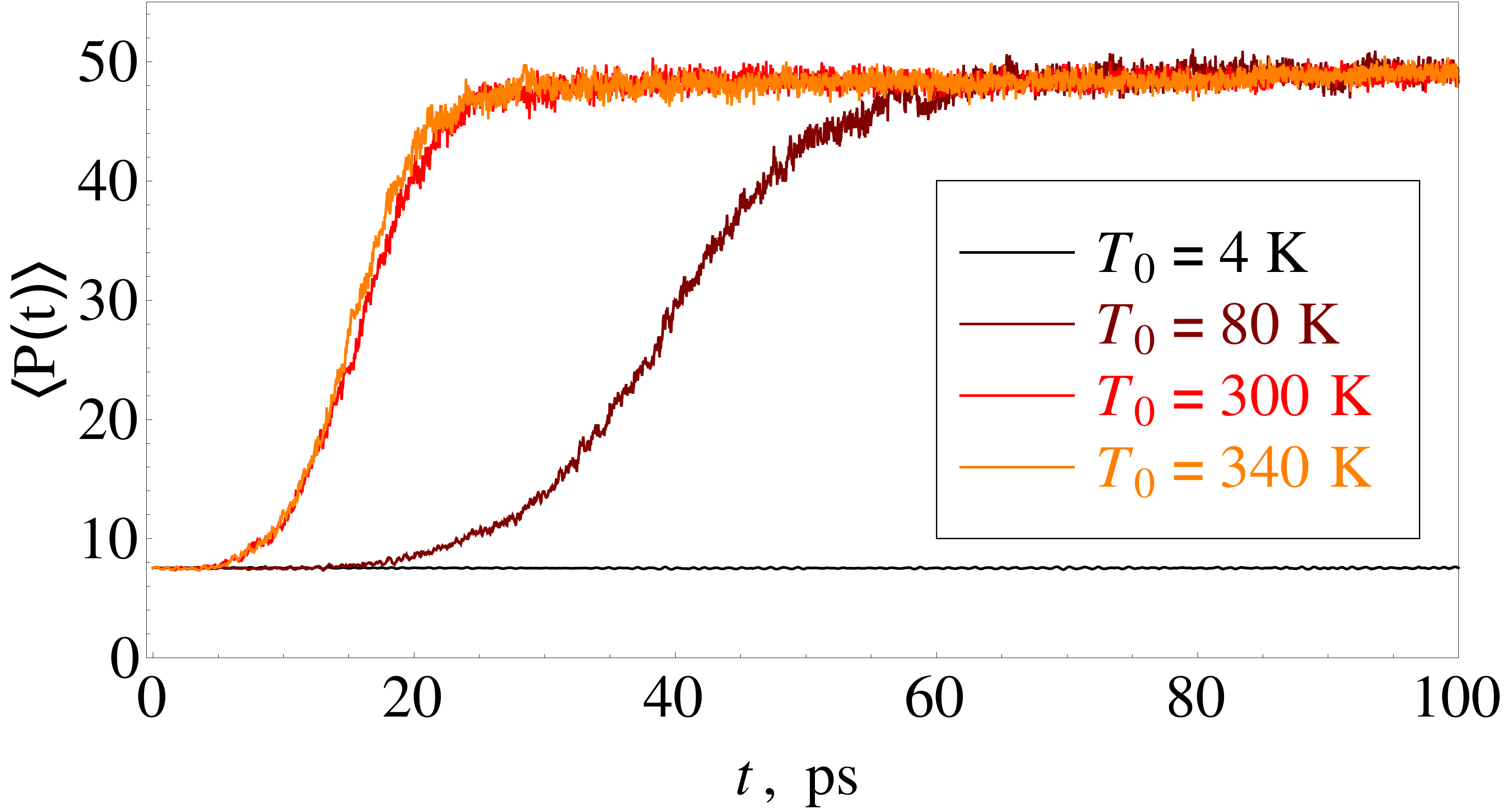}
\caption{
Time evolution of the ensemble averaged PN for different bath temperatures, using the
polaronic solution as the initial condition.  For each temperature, 100
trajectories where used to calculate the ensemble average
$\langle\mathrm{PN}\rangle\equiv \langle\mathrm{P}(t)\rangle$.
}
\label{fig:PNevolutionInitialPolaron}
\end{figure}

The ensemble averaged time evolution of the PN for different bath temperatures $T_0$ is shown in Fig.~\ref{fig:PNevolutionInitialPolaron} which demonstrates that the polaron lifetime becomes shorter for higher temperatures, as can be expected.

\begin{figure}[b]
\includegraphics[width=\linewidth]{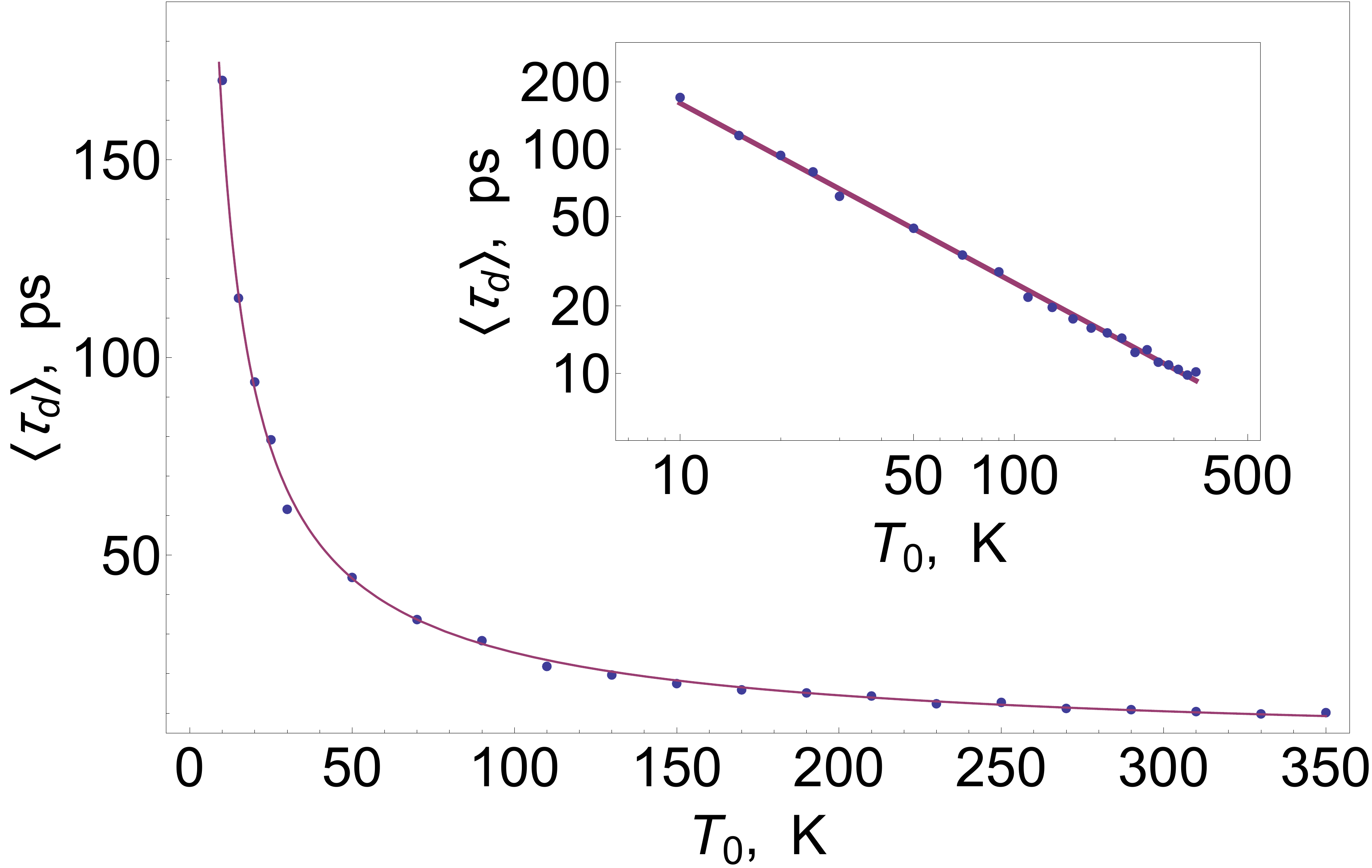}
\caption{
Ensemble average polaron lifetime $\langle\tau_d\rangle$ as a function of the bath temperature $T_0$. Dots represent numerical results while solid line are best power law fits to the numerical data. It can be seen in the log-log inset that the data follows the power law; the exponent giving the best fit is $p \approx -0.8$.
}
\label{fig:delocTimeInitialPolaron}
\end{figure}

Using the above mentioned definition of the polaron lifetime, we calculated its dependence on the bath temperature $T_0$, as shown in Fig.~\ref{fig:delocTimeInitialPolaron}.  This dependence turns up to be a power law one with the exponent $p \approx -0.8$ (see the inset demonstrating the best power law fit to the data).

\begin{figure}[h!]
\includegraphics[width=\linewidth]{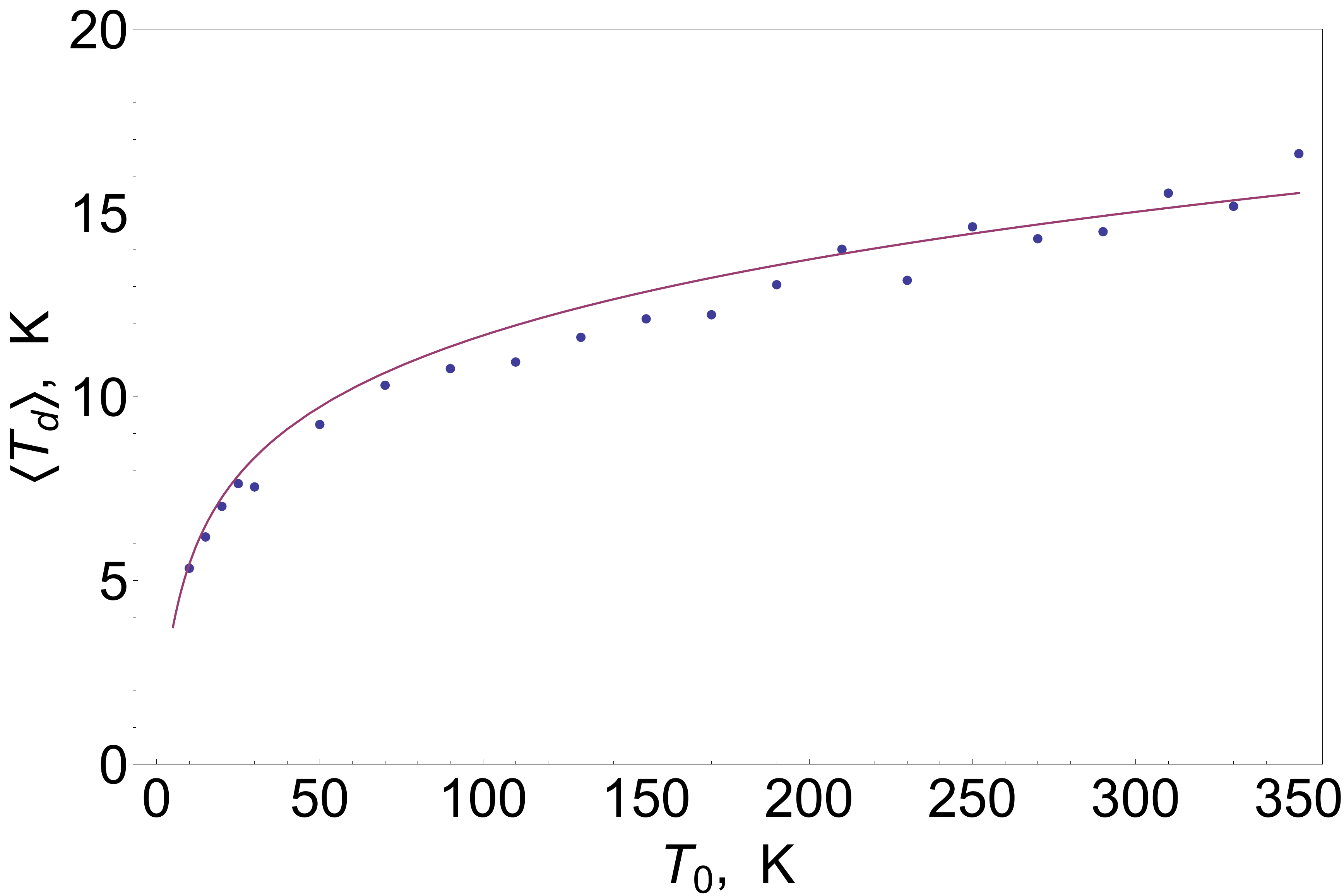}
\caption{
Ensemble averaged temperature $\langle T_d\rangle$ (temperature at the time $\langle\tau_d\rangle$) as a function of the bath temperature $T_0$ --- dots, and its theoretical estimation $T(\tau_d)$, using the best fits from  Figs.~\ref{fig:tauDetermination}, \ref{fig:delocTimeInitialPolaron} 
and Eq. (\ref{eq:thermprocess}) --- the solid curve.
}
\label{fig:delocTemperatureInitialPolaron}
\end{figure}

Using the thermalization time scale $\tau_0$ and the obtained polaron lifetime power law dependence on the bath temperature $T_0$, it is straightforward to calculate the chain temperature $T_d$ at $t=\tau_d$, {\it i.  e.}, at the moment when the polaron breaks up and the electron starts extending over the whole chain.  The result is shown in Fig.~\ref{fig:delocTemperatureInitialPolaron}, where the points represent the mean temperature calculated for all considered trajectories using the kinetic energy at $t=\tau_d$ as a function of the temperature $T_0$.  The solid line shows the theoretical estimation of $T_d=T(\tau_d)$ according to Eq.~(\ref{eq:thermprocess}) in which the best fit parameters from the previous calculations are used.  The two results are in good agreement with each other, which indicates that our considerations are self consistent. Our findings suggest that the polaron breaks up for not so low temperatures ($T_0>10$K). Moreover, it breaks up long before the target chain thermalization temperature $T_0$  is reached because $T_d < T_0$ and $\tau_d < \tau_0$ for all considered cases. We conclude therefore that that the polaronic configuration can hardly be expected to be a relevant state of the system in the studied temperature range within the framework of the PBH model. We comment in detail on the low temperature regime dynamics in Sec.~\ref{noPBH}.



We point out that in the current section we considered a very artificial scenario in which a polaron existed initially. Nevertheless, we found that polarons are not stable even under such favorable conditions. Below we address a more realistic case.

\section{Dynamics of an electron in a thermalized lattice
\label{sectionElectron}}

In this section we address the dynamics of a charge injected in a thermalized chain which has the temperature of the heat bath: $T=T_0$. Such an initial condition can occur when an electron belonging to the chain environment hops to one of the nucleotides. If the polaronic solution is relevant and stable one would expect that the system would evolve towards it.

To simulate this scenario we proceed as follows. We solve the Langevin equation for the molecule chain until it reaches thermal equilibrium at the bath temperature $T_0$.  Then, at $t=0$ an electron is created at one of the sites (hereafter, the site number $N/2$) in the form of the $\delta$-function wave packet and the dynamics of the coupled system is studied. Below we address such dynamics for different temperatures and argue that there can be very different qualitative regimes of the charge density evolution. To this end, we study both the participation number characterizing the spacial extent of the wave function as well as particular trajectories in order to get insight into the underlying mechanisms of the electronic density dynamics.

\begin{figure}[h]
\includegraphics[width=\linewidth]{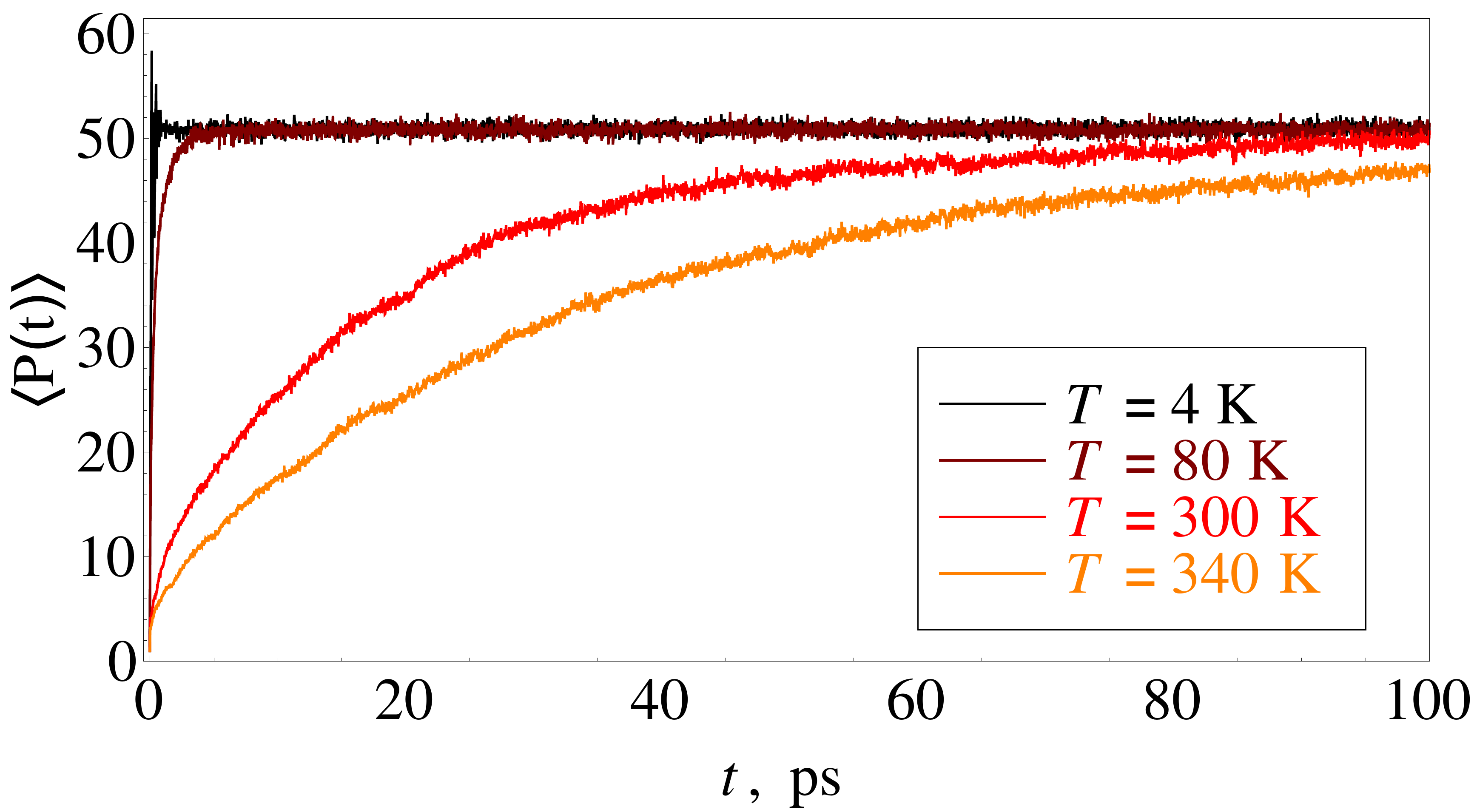}
\caption{Evolution of the average participation number $\langle\mathrm{PN}\rangle=\langle\mathrm{P}(t)\rangle$ for different temperatures, 
using an electron injected at $n=50$ in a previously thermalized chain as the initial 
configuration. For every temperature, 100 paths where used to calculate the ensemble average. }
\label{fig:PNevolutionElectronInjected}
\end{figure}

First, we consider the ensemble averaged time dependence of the participation number for different temperatures. The results are presented in Fig. \ref{fig:PNevolutionElectronInjected}. The figure shows no traces of the polaron formation for all temperatures: the PN grows to its maximum value corresponding to the charge density distributed almost homogeneously over the whole sample. As can be seen from the figure, the wave function expands faster for lower temperatures, which may seem counterintuitive. Indeed,
the trend is just the opposite compared to that observed in the previous section where we showed that the polaron lifetime (delocalization time) is larger for lower temperatures. 

To explain the increase of the wave function expansion time we note that because of the 
structure of the interaction term in (\ref{eq:Hinter}) lattice vibrations result in a random time-dependent potential for an electron. Due to strong asymmetry of the Morse potential, lattice distortions are predominantly positive ($y_n>0$). The latter results in appearance of positive values of the on-site potential [the diagonal of the electronic Hamiltonian in Eq.~(\ref{eq:Hlat})], which act as barriers for the wave function. The amplitude of these barriers can be characterized by the temperature dependent typical value of the stretchings. Such a value increases with the temperature, making the system more ``disordered" for the electron, which slows down the electron dynamics. 

The above over-simplified qualitative argument is based on the integrated characteristics of the system, the PN. Therefore, such a consideration can and actually does overlook important details of the charge density dynamics. To get an insight into the latter we proceed by studying particular typical trajectories for different temperatures, for which purpose we consider relatively short chains ($N=100$) in order to be able to see clearly important features of the system dynamics from various map plots (see below).

\subsection{Very low temperatures: ballistic propagation}

\begin{figure}[h]
\includegraphics[width=\linewidth]{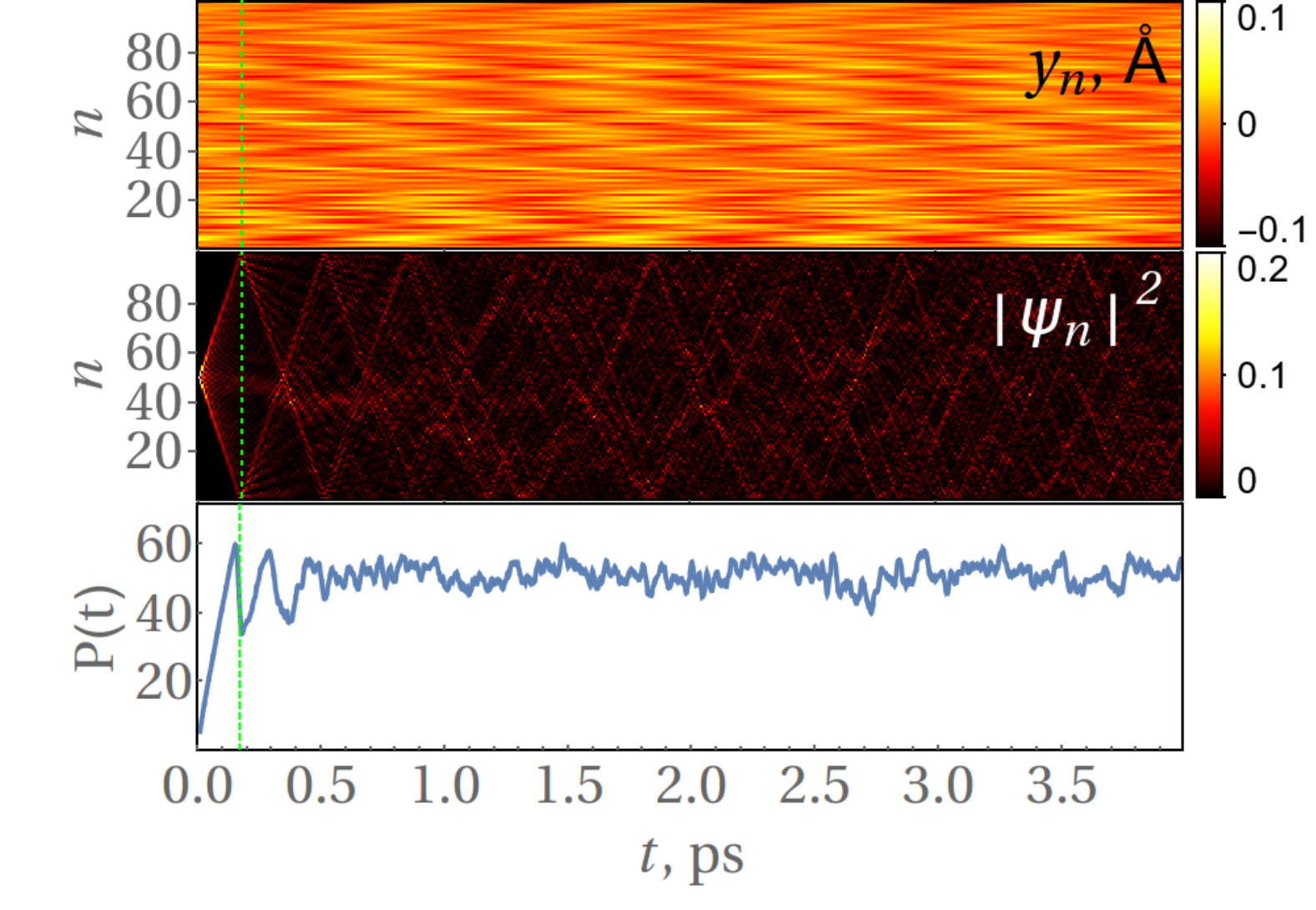}
\caption{
Density plots of the time evolution of the bond stretchings $y_n$ (upper panel) and the electron probability density $|\psi_n|^2$ (middle panel). The lower panel shows the time dependence of the participation number. All results are calculated for the case when an electron is injected into a thermalized chain at $t=0$, vertical green line gives the time of flight (see text for details). The chain temperature is $T=4$ K.  
}
\label{fig:elInjEvolution4}
\end{figure}

We start with the case of a very low temperature: $T=4$K. The results are presented in Fig.~\ref{fig:elInjEvolution4}. The upper panel shows the time evolution of the bond stretchings $y_n$, the middle one -- the probability density $|\psi_n|^2$, while the lower one -- the participation number PN. The upper panel demonstrates that the typical bond stretching is very small: on the order of $0.05\mbox{\AA} \ll 1/\alpha$. In this case, the Morse potential can be approximated by the parabolic one:
\begin{equation}
V(y_n)
\simeq V_0 \,\alpha^2 y_n^2\ ,
\end{equation}
which has the period of oscillation $2\pi\sqrt{\mu/2V_0\alpha^2}$ of about $0.88$ ps. Patterns with this period are clearly observed in the upper panel of the figure. Note also the evident absence of the influence of the electronic density on the lattice dynamics: it is qualitatively the same during the initial phase when the electron is localized and in later phases when the electronic density is extended over the whole chain. Besides, the whole displacement pattern is very homogeneous showing no outstanding features (below we will contrast this very uniform pattern to those characterized by large fluctuations at higher temperatures).

The middle panel of Fig.~\ref{fig:elInjEvolution4} shows typical features of the ballistic propagation for $t<0.5$ ps: the clearly straight tracks crossing the whole sample. The existence of such a pattern can be explained as follows. Using the dispersion relation for a one-dimensional ideal chain, we can calculate the group velocity:
\begin{equation}
v_g(k)= 
 \frac{ 2\,a\,J}{\hbar} \sin\left(\frac{\pi\,k}{N+1}\right),\; k =1,2,\ldots N
\end{equation}
where $a$ is the chain lattice constant and $k$ is the state number. The initial $\delta$-function wave packet projects almost uniformly over the eigenstates with odd values of $k$. The fastest eigenmodes belong to the center of the energy band
($k\approx N/2$); they have approximate group velocity $2aJ/\hbar$ of about $300\,a$/ps, and therefore travel over a half of the chain in about $0.17$ ps (the vertical green line in Fig.\ \ref{fig:elInjEvolution4} gives this time scale). Note that the latter time of flight is about six times smaller than the characteristic oscillation period ($0.88$ ps), so for these states the electron-phonon coupling reduces to interaction with a virtually static disorder. These faster states are less affected by disorder than slower edge states ($k\sim 1$ or $k \approx N$) because of the motion averaging effect: the faster a state is propagating the better it is averaging the disorder, making the latter effectively weaker. Also, the PN is growing very fast which reduces the typical wave function amplitude and consequently the electron-phonon coupling even further~\cite{Malyshev1998}. Therefore, the part of the wave packet corresponding to the faster band-center states travels ballistically during the first phase of the propagation. When it reaches the hard boundary of the chain it gets reflected and travels backwards, which can clearly be seen in the Fig.~\ref{fig:elInjEvolution4}. The ballistic propagation manifests itself also in the characteristic interference fringes of the PN (see the lower panel of the figure at $t<0.5$ ps). 

\begin{figure}[t]
\includegraphics[width=\linewidth]{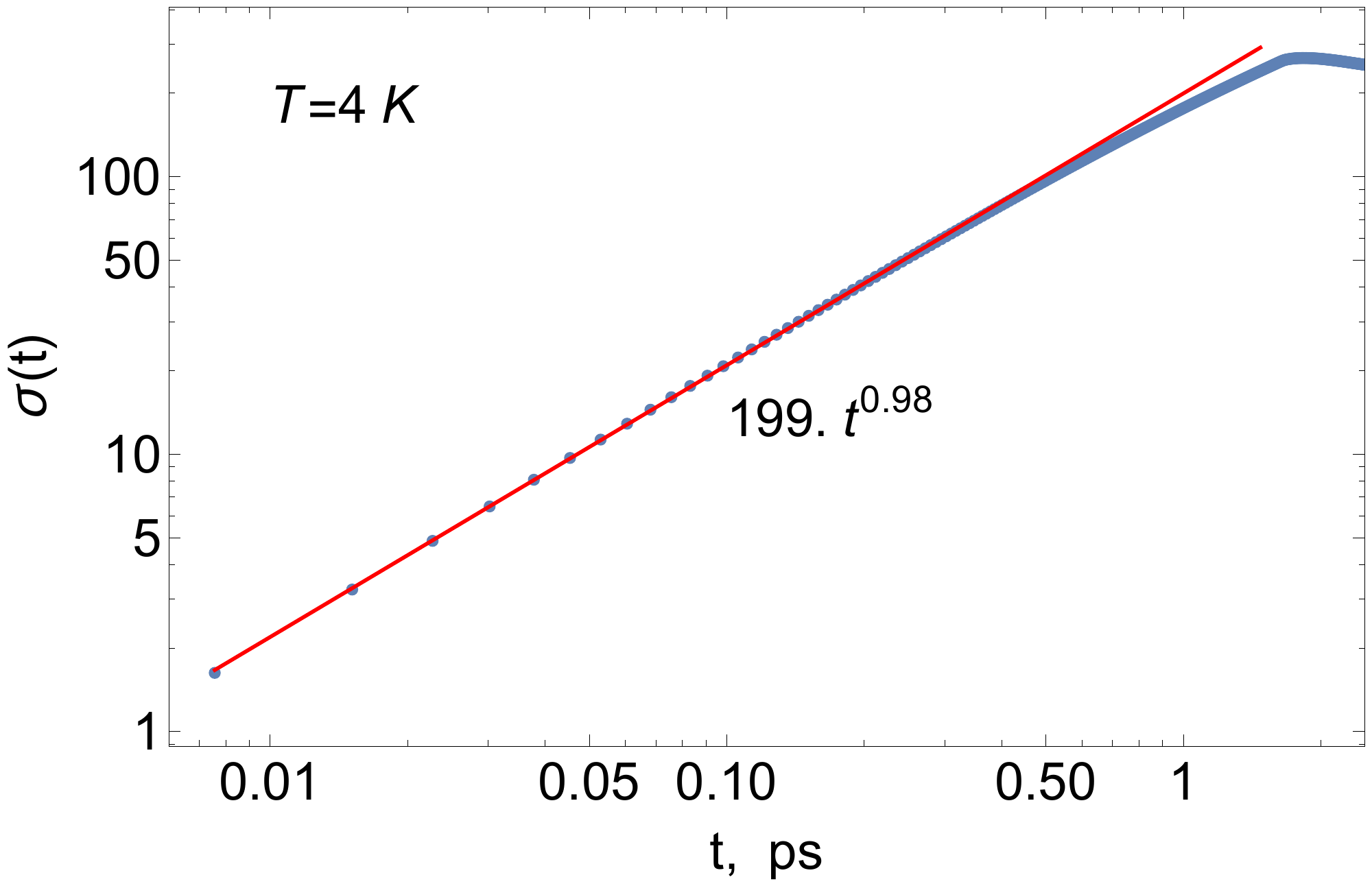}
\caption{
The dependence of the ensemble averaged standard deviation of the electronic density (packet width) on time calculated for a chain of $N=1000$ sites at $T=4$ K.  Solid red line gives the best power law fit to the data within the region $t<0.3$ ps. The best fit function is shown next to the line.
}
\label{fig:sigma4K}
\end{figure}

Even in the current case of very low temperatures, when the oscillation period is the shortest possible (about $0.88$ ps), all characteristic lattice timescales are much larger than that of the electronic subsystem: the average group velocity $\bar{v}_g$ is on the order of $200\,a$/ps, so the corresponding timescale $a/\bar{v}_g\sim 0.005$ ps. Thus, the lattice is unable to adapt itself to fast changes of the electronic density and, for example, a polaron can not form. Besides, amplitudes of the lattice vibration are small and very homogeneous, so they do not provide high potential barriers for the charge carrier, which could scatter it back efficiently as in the case of higher temperatures (see below).  

In order to analyze electron dynamics and compare to standard transport scenarios, hereafter, we also calculate the time evolution of the wave packet width, {\it i. e.}, the standard deviation of the electronic density for a given trajectory, defined as follows:
\begin{equation}
s(t)=\sum_n(n-x(t))^2\,|\Psi_n(t)|^2\ ,
\end{equation}
where $x(t)$ is the wave packet centroid position:
\begin{equation}
x(t)=\sum_n n\,|\Psi_n(t)|^2\ .
\end{equation}

Figure \ref{fig:sigma4K} shows the dependence of the ensemble averaged packet width $\sigma(t)=\langle s(t)\rangle$ on time calculated for a chain of $N=1000$ sites at $T=4$ K. The packet width saturates at long times when the packet expands over the whole chain - this is just a finite size effect. The best power law fit, given by the red line in the figure, has the exponent very close to unity. Thus, we conclude that, within the framework of the considered model, in the regime of very low temperatures a substantial part of an initially stationary electron wave packet can expand ballistically over long distances, on the order of a hundred of lattice constants. Similar results were reported recently in Ref.~\onlinecite{DPC16} where ballistic propagation of Frenkel excitons in molecular aggregate systems was discussed.

\subsection{Intermediate temperatures: diffusive regime}

\begin{figure}[ht]
\includegraphics[width=\linewidth]{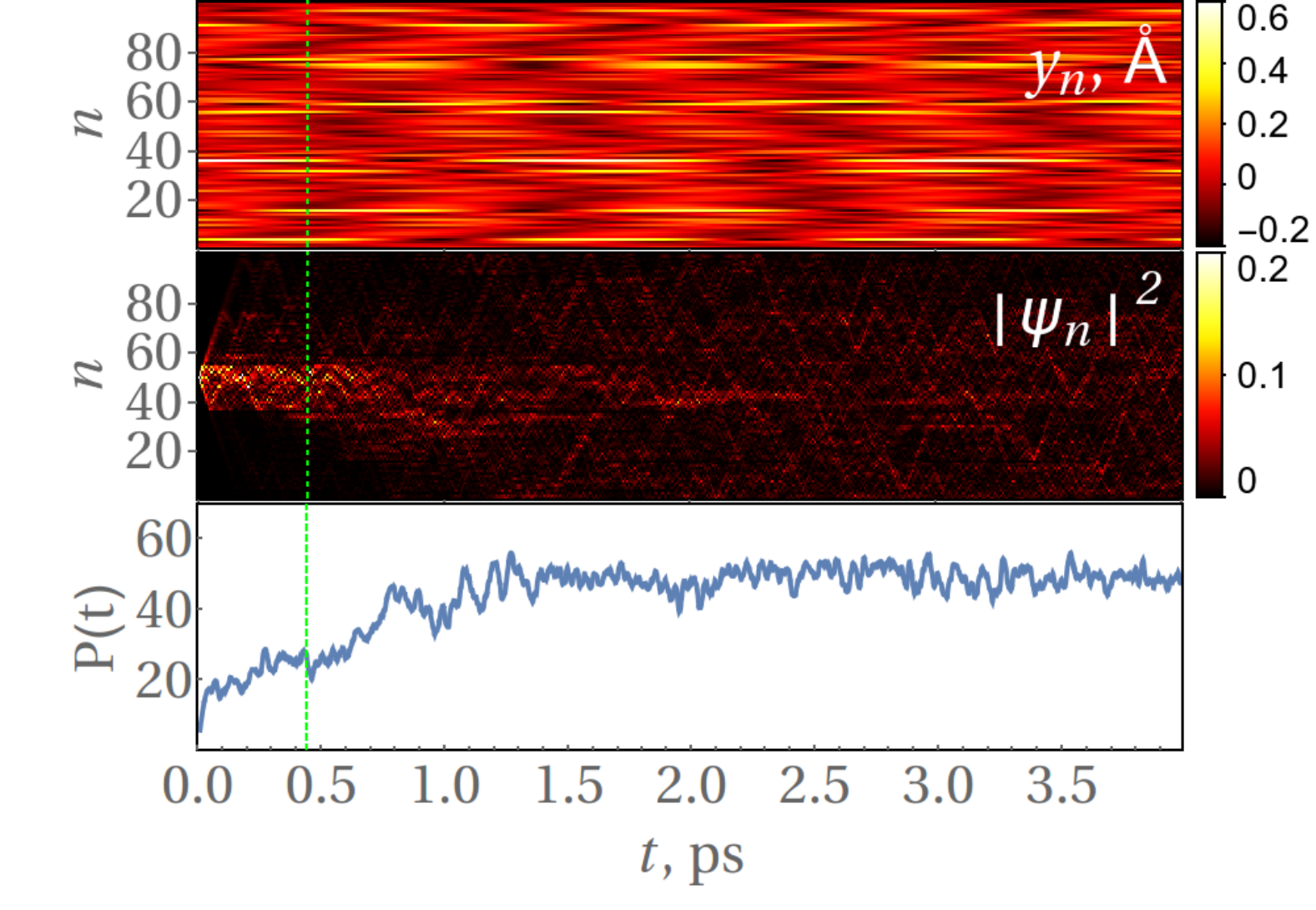}
\caption{
Same as in Fig.~\ref{fig:elInjEvolution4} but for $T=80$K.  
}
\label{fig:elInjEvolution80}
\end{figure}

Next, we consider the system dynamics in the regime of low to intermediate temperatures. Results of our calculation of one particular system trajectory are presented in Fig.\ \ref{fig:elInjEvolution80} which has the same layout as Fig.\ \ref{fig:elInjEvolution4}. As in the previous case, the upper plot of the bond stretchings shows clear traces of oscillatory motion but the typical value of the stretching now is about $0.2\mbox{\AA}$. In this case, the Morse potential is not quadratic anymore and the characteristic period is somewhat larger than that of the harmonic potential approximation used above. More importantly, the pattern is not as homogeneous as before: there are clearly visible long-living stronger fluctuations (see the stripes with light yellow parts marking large positive stretchings). For example, see the strong oscillatory fluctuation around $n=70$: its period is somewhat larger than the harmonic one (we mark the harmonic half period of $0.44$ ps by the vertical green line in the figure for a reference). When this oscillating chunk of the lattice is stretched it represents a sufficiently high potential barrier for an electron due to the interaction term Eq.~(\ref{eq:Hinter}). Note how a part of the wave packet is reflected back by that potential barrier (see the middle panel) and by another one at about $n=35\div 40$. So, while these barriers are high the charge is largely confined between them. However, during the other part of the oscillation period bond stretchings shrink to about zero and barriers disappear letting the charge density to leak out and propagate further. 

\begin{figure}[h]
\includegraphics[width=\linewidth]{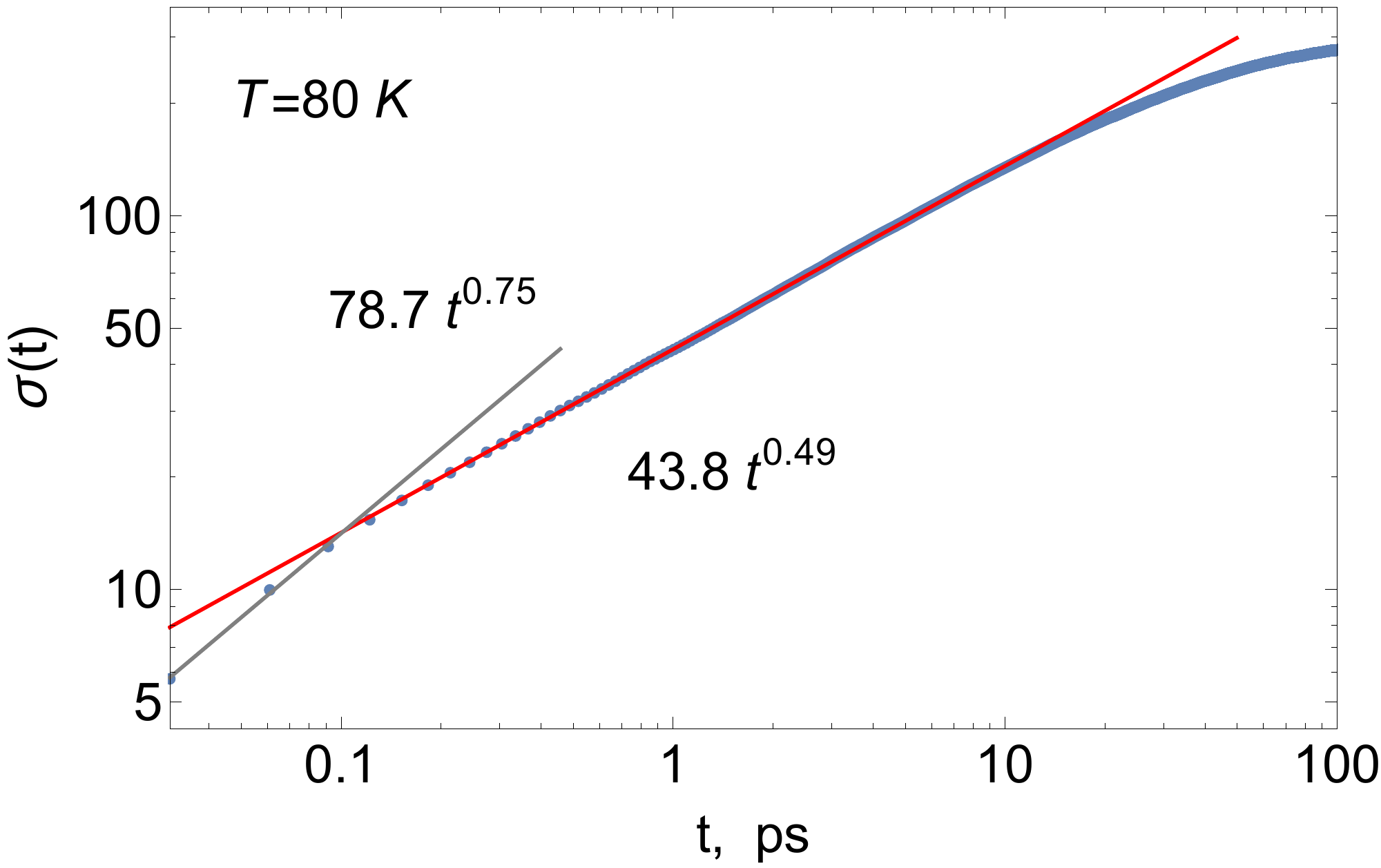}
\caption{
Same as in Fig.~\ref{fig:sigma4K} but for $T=80$ K. Here, solid red line gives the best power law fit to the data within the region $0.1<t<10$ ps, while the grey line -- in the region $t<0.1$ ps.
}
\label{fig:sigma80K}
\end{figure}

Therefore, the chain oscillation period sets a natural time scale for the charge propagation. It is important that the system is far from the phase transition point at such temperatures and therefore it can be characterized by finite size and time scales of oscillations of either base pairs or finite chunks of them. These oscillations rise and lower potential barriers for an electron; the phases of the oscillations are random, so the charge migration mode is very similar to that of a random walk resulting eventually in diffusive regime of the charge dynamics, which can be seen in the Fig.~\ref{fig:sigma80K}.

\subsection{On the PBH model for temperatures below the environment freezing point.\label{noPBH}}

We have been considering the system dynamics at relatively low temperatures of the environment forming the thermal bath. In the case of the DNA and other organic molecules this environment is typically water based, so we have been studying these systems for temperatures well below the freezing point of the medium they are embedded into. There are several concerns related to the latter circumstance we would like to discuss here.

For temperatures close to the medium melting point (usually about $0$ C) the typical stretching is on the order of an $\mbox{\AA}$ (results not shown here). Recall that in the case of the DNA stretchings $y_n$ are elongations of H-bonds between the base molecules forming the two DNA strands. Thus, elongations represent relative displacement of large base molecules. If the environment is a fluid and fast enough, it can easily adapt itself to relatively slow changes in base pair configuration. However, below the freezing point the medium is glassy and it can hardly support such large displacements of bulky objects embedded into it. Therefore the usability of the PBH model for such temperatures can be questioned. 

Unphysically large displacements do not occur at low temperatures. See, for example, the upper panel of Fig.~\ref{fig:elInjEvolution80} from which the characteristic bond stretching at $T=80$ K can be estimated to be on the order of $0.2\mbox{\AA}$. However, there is another concern: the stochastic term in the Langevin equation, which models the action of the bath is uncorrelated. This is reasonable for a very fast liquid environment but does not seem to be adequate for a glassy matrix when the local configuration of each chain site is random but frozen. Such random local environment introduces a static disorder into the system, resulting in characteristic energy structure of localized states (see, e. g. Refs.~\onlinecite{Malyshev1991a,Malyshev1991b,Malyshev1993,Malyshev1995,Malyshev2001a,Malyshev2001b}), while the action of the bath can be considered as the interaction with phonons of the glassy host. We believe that such a model is more realistic for the considered systems at low temperatures. We note finally that the model of phonon-assisted transport has been used very successfully for describing quantitatively various properties of molecular aggregates at low temperatures (see, for example, Ref.~\onlinecite{malyshev2003a,malyshev2003b,malyshev2007b} and references therein) as well as some aspect of charge transport in the DNA\cite{Malyshev2009}.

In the next section we address the system dynamics at temperatures above the environment melting point, when the PBH model is more appropriate.

\subsection{Room temperature; mixed regime}

\begin{figure}[ht]
\includegraphics[width=\linewidth]{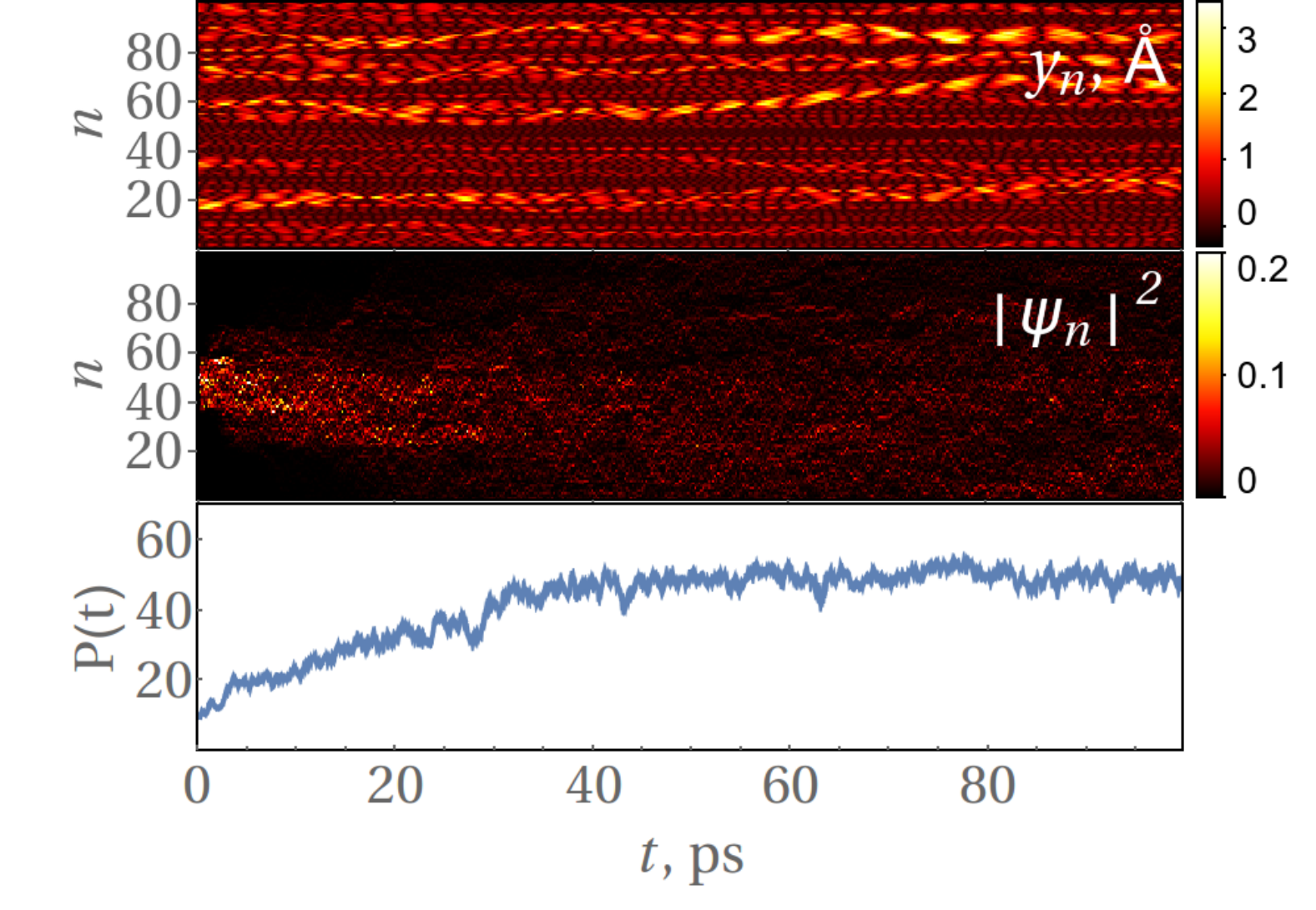}
\caption{
Same as in Fig.~\ref{fig:elInjEvolution4} but for $T=300$K.  
}
\label{fig:elInjEvolution300}
\end{figure}

In this section we study the case of high temperatures: above the medium melting point but still well below the critical one. Results of our calculation of one particular typical system trajectory for $T=300$ K are presented in Fig.\ \ref{fig:elInjEvolution300}. One can see from the upper panel that the characteristic displacements are on the order of an $\mbox{\AA}$, in which case the interaction potential $W(y_n,y_{n+1})$ (which is negligible for small stretchings) starts playing an important role. This manifests itself in the appearance of specific strong fluctuations: chunks of several sites moving in a correlated way (for example, in phase). These chunks are substantially heavier than a single base, so their dynamics is slower and, as the panel suggests, they can be very long living. See, for example the brighter yellow traces showing such fluctuations around $n=20\div 30$ and $n=60\div 70$. Their characteristic period is on the order of several ps, up to about $5$ ps. Note also that the part of the figure enclosed between them is considerably darker which corresponds to smaller stretchings. This means that with respect to that enclosed part the two stronger fluctuations act as much higher potential barriers for electron wave packet. Indeed, from the electron density dynamics (see the middle panel) one can see that the wave packet is visibly confined between the two trains of oscillating high barriers for a very long time: until about $t=20\div 30$ ps. These features are also reflected in the behavior of the PN (see the lower panel) and the  dynamics of the packet width (as we discuss below). 

\begin{figure}[h]
\includegraphics[width=\linewidth]{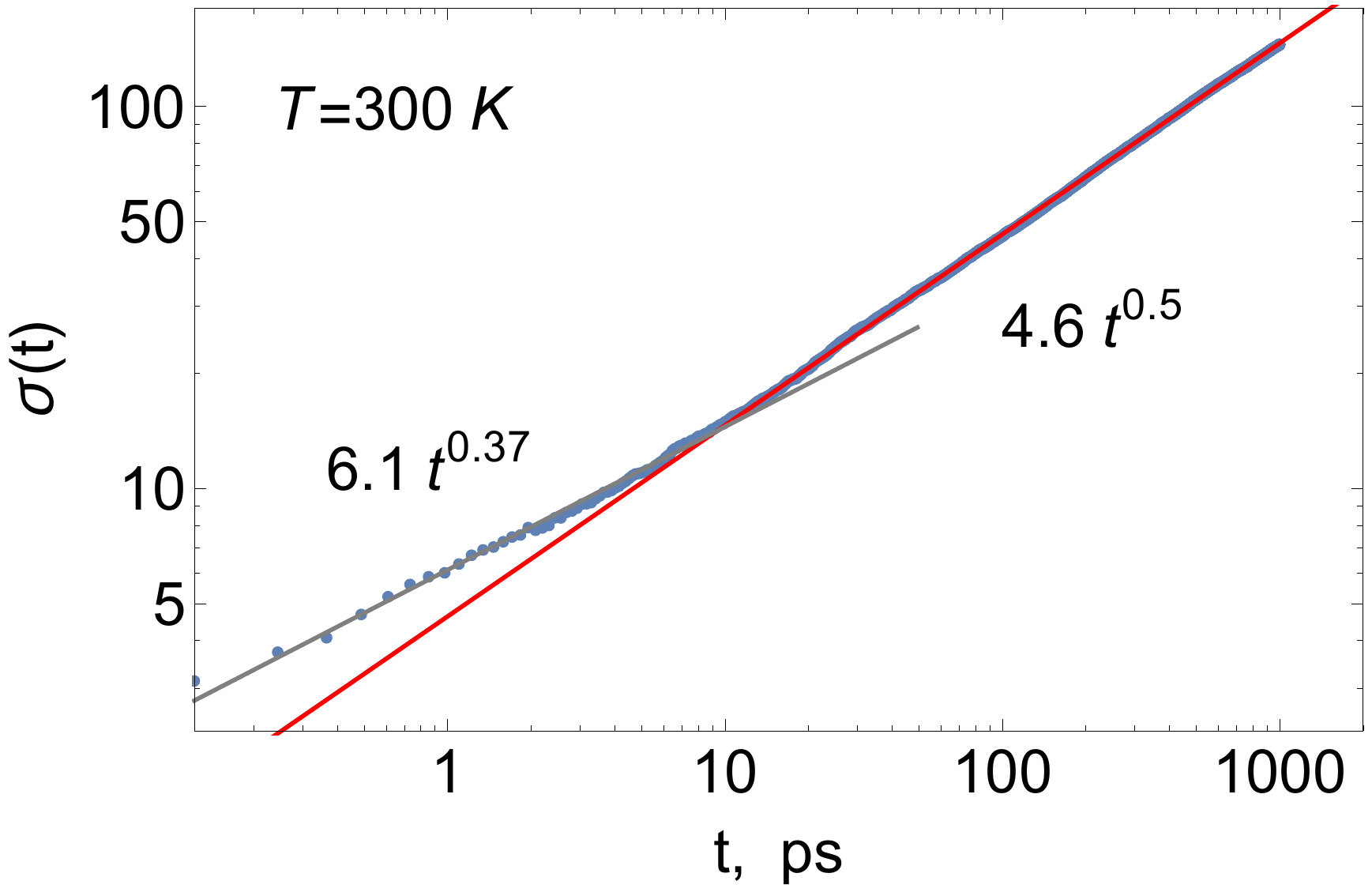}
\caption{
Same as in Fig.~\ref{fig:sigma4K} but for $T=300$ K. Here, solid red line gives the best power law fit to the data within the region $t>10$ ps, while the grey line -- in the region $t<10$ ps.
}
\label{fig:sigma300K}
\end{figure}

Figure \ref{fig:sigma300K} shows the dependence of the ensemble averaged packet width on time calculated for a chain of $N=1000$ sites at $T=300$ K. Two different transport regimes can be distinguished from the figure: for $t<10$ ps and $t>10$ ps. Gray and red solid lines give the best power law fits to the numerical data within the two time ranges. As expected, the packet expansion is diffusive in the long time limit (with the exponent $0.5$). At this temperature the fluctuations governing the charge migration have characteristic space and time scales, which set natural time scales for the random walk process. From the upper panel of Fig.\ \ref{fig:elInjEvolution300} the following estimations can be made: the typical spacing between adjacent {\it strong} fluctuations is on the order of $10$ lattice constants and their period is on the order of several ps. Only when the propagation time substantially exceeds such a time scale, the diffusive nature of the propagation can manifest itself. The latter agrees well with $t=10$ ps being the time separating the two regimes in this case. On the other hand, strong fluctuations split the chain into chunks which appear to act as new super-sites. The size of the latter gives the characteristic length scale of the random walk on a chain of such super-sites. This is also in agreement with the fact that the diffusive spreading starts when $\sigma(t)>10$ (which happens at $t=10$ ps). We can argue therefore that the diffusive regime is governed by space and time characteristics of {\it strong} fluctuations.

For shorter times ($t<10$ ps) the packet dynamics is clearly sub-diffusive with the exponent of about $0.37$. The latter can be explained by the fact that the wave packet is efficiently confined by neighboring strong and long-living fluctuations (as we have discussed above), which traps the charge density within a limited chunk of the whole DNA chain for substantially long time, resulting in a slowing down of the overall packet dynamics and making it sub-diffusive within this time range. We can conclude that the charge migration in DNA at room temperature is sub-diffusive up to about $10$ ps during which a wave packets spreads over about $20$ base pairs. 

We checked that the lattice dynamics does not change if the vibrational and the electronic parts of the system are decoupled (results not shown here) suggesting that the influence of the charge dynamics on the vibrational one is negligible. 
However, the electron-phonon interaction is determining for charge dynamics in the case of high temperatures because charge migration is governed by the lattice fluctuations. Such a mechanism can be called a fluctuation-bound charge migration.

\subsection{Close to critical temperatures; sub-diffusive regime}

\begin{figure}[t]
\includegraphics[width=\linewidth]{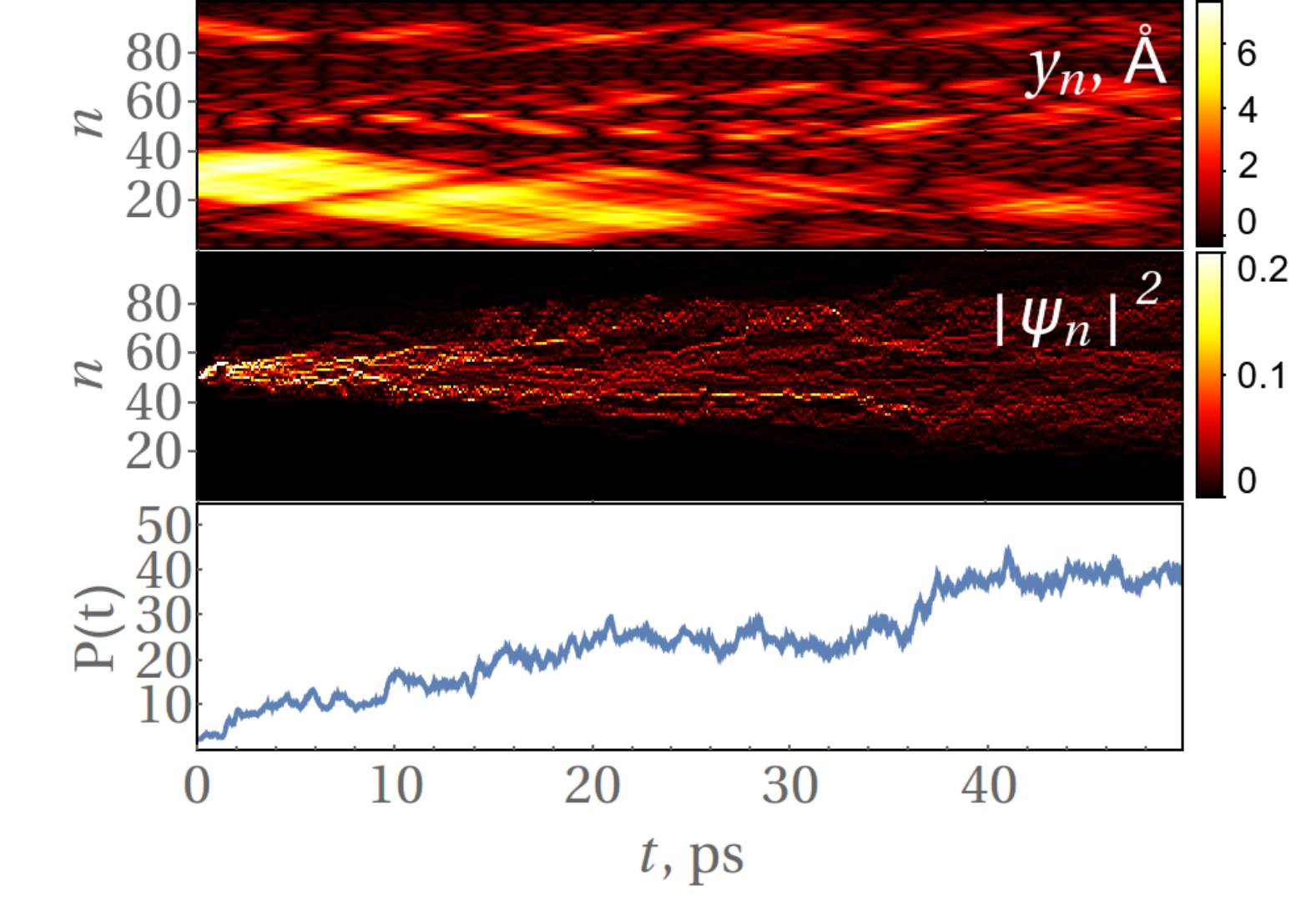}
\caption{
Same as in Fig.~\ref{fig:elInjEvolution4} but for $T=340$K.  
}
\label{fig:elInjEvolution340}
\end{figure}

In this section we address briefly a very special case in which the system is close to its phase transition point. Within the considered model the melting transition temperature is about $356$ K~\cite{Peyrard1989,Dauxois1993}. Results of our calculation for one particular system trajectory at $T=340$ K are presented in Fig.\ \ref{fig:elInjEvolution340}. It is known that in the vicinity of a phase transition point fluctuations become critical or scale-free, which means that both small and very large (diverging at the transition) fluctuations can co-exist. In the case of the DNA such large fluctuations correspond to so called bubbles --- large openings in the double helix chain --- resulting finally in the DNA denaturation or melting, {\it i. e.} complete separation of the two strands. In the PBH model the melting corresponds to divergence of the H-bond stretching $y_n$. The upper panel of the figure demonstrates fluctuations of very different sizes --- a precursor of the scale-free or critical ones. They can be very long living and therefore, as we have argued above, the wave packet can be confined efficiently by them for considerable amounts of time.  The result of such strong confinement can be clearly seen in the middle and the lower panel of the figure. Note, in particular, that the PN has flat plateaus, for example, for $5\lesssim t \lesssim 10$ or $20\lesssim t \lesssim 35$ corresponding to time windows when the wave packet is localized between a pair of bubbles. The characteristic space and time scales of the charge dynamics are determined by those of the chain fluctuations. Due to the criticality of fluctuations at the transition point they are scale free both in space and time and therefore the charge dynamics can be expected to manifest anomalous behavior. We observed the sub-diffusive wave packet dynamics with $\sigma(t)\sim t^{0.48}$ for the critical temperature $T=356K$. However, a direct study of such a dynamics in more detail would require very large size and time scale calculations. An alternative way to detect anomalous transport properties is by studying fluctuations. For example, one can address the relative fluctuation of the wave packet size. At the transition point, both the size and its fluctuation is expected to grow at the same rate in the thermodynamic limit, so their ratio should tend to a constant (see, e. g., Ref.~\onlinecite{Malyshev2004} and references therein). Below we use the latter approach to support our point.

\begin{figure}[h]
\includegraphics[width=\linewidth]{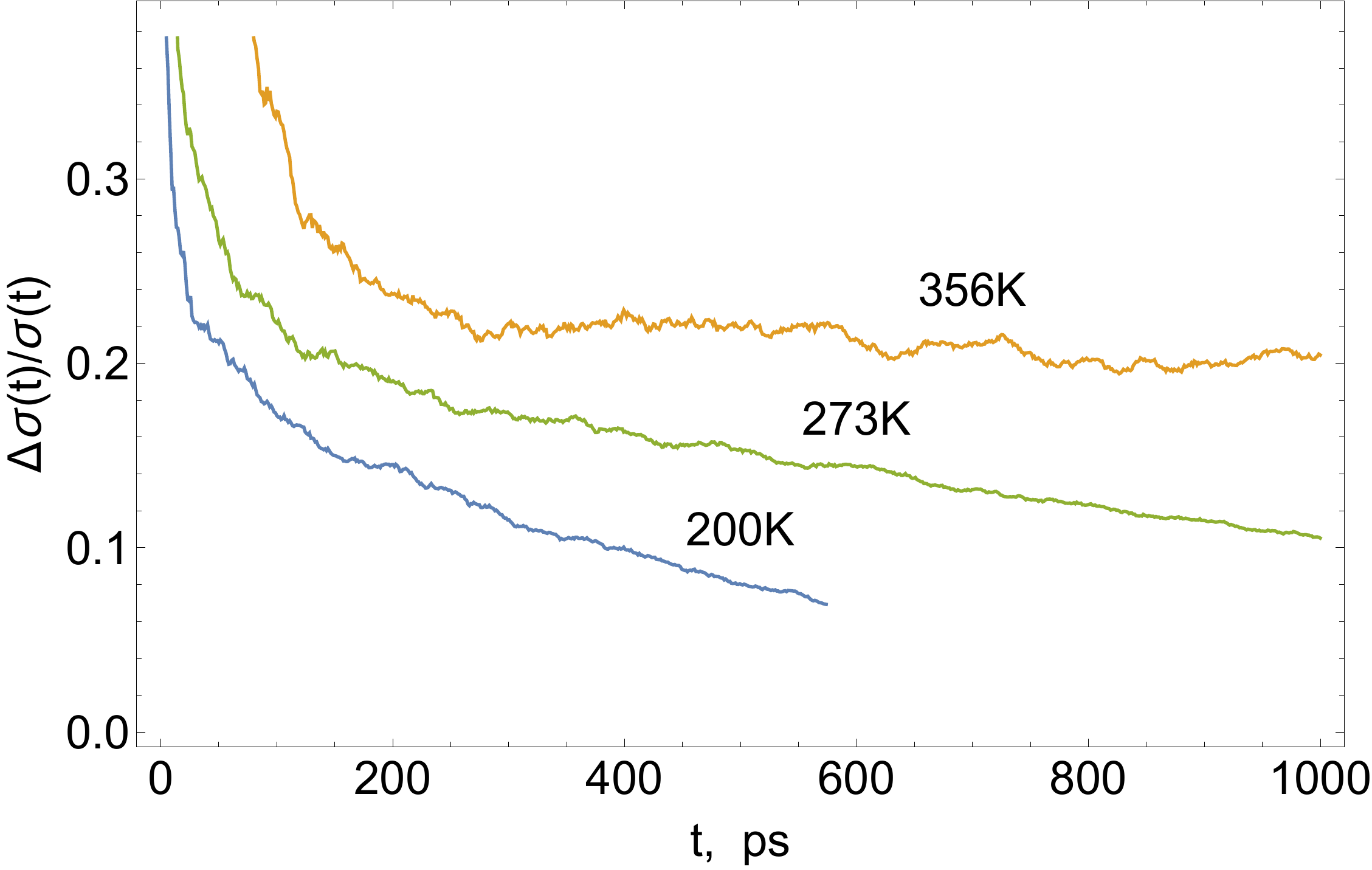}
\caption{
Relative error of the wave packet width: the ratio of the standard deviation $\Delta\sigma(t)$ to the width $\sigma(t)$ for different temperatures (the system size is $N=1000$, in all cases $\sigma(t)$ is considerably smaller than $N$ so size effects are negligible).
}
\label{fig:reller}
\end{figure}

To illustrate the emergence of the anomalous regime of the charge migration at the transition temperature we plot in Fig.~\ref{fig:reller} the evolution of the relative fluctuation: the ratio of standard deviation of the electronic wave packet size $\Delta\sigma(t)$ to the size $\sigma(t)$ itself, for different temperatures. For temperatures well below the critical one ($T=200K$ and $T=273K$ in the figure) the relative error is decaying with time, which reflects the fact that the role of the fluctuations decreases and the expanding wave packet has a well defined size. Contrary to that, at the critical temperature, $T=356K$, the relative error tends to a constant, indicating that the fluctuations of the wave packet size remain on the order the size itself in the thermodynamic limit --- a typical feature of anomalous transport.~\cite{Malyshev2004} A more detailed analysis of this case goes far beyond the scope of the present work.
%

\section{Conclusions}

We have studied various charge migration mechanisms in the DNA molecules within the framework of the Peyrard-Bishop-Holstein model which considers coupled dynamics of the electronic and vibrational degrees of freedom of the system. At zero temperature the minimum energy configuration of the model is polaronic. For this reason, first, we analyzed the polaron dynamics during the DNA chain thermalization, connecting the system with a polaron as the initial condition at $T=0$ to a heat bath with a finite temperature $T_0$. We found that for the whole range of studied temperatures ($T_0\approx 10 \div 356$ K) the polaron has a finite lifetime and, more importantly, it brakes up long before the vibrational subsystem reaches the equilibrium at the temperature of the bath. We believe that the result suggests that the relevance of the polaron charge migration mechanism within the framework of the model can be questioned because a polaron does not seem to be a stable configuration at finite temperatures.

Second, we studied the system dynamics under a more realistic initial condition: a charge injected into a thermalized DNA chain. We found no traces of polaron formation either. Rather, we observed and analyzed a variety of transport regimes at different temperature ranges. At low temperatures the charge transport showed up to be quasi-ballistic: a wave packet can expand almost ballistically over several dozens of system sites (base pairs). We argued, that for higher temperatures that are below the environment (bath) freezing point the applicability of the PBH model can be questionable due to large typical values of mechanical displacements in the vibrational subsystem, which can hardly be realistic in a glassy host medium. 

Finally, we found that above the environment melting point, which is the most interesting temperature range for real biological systems where the PBH is probably an appropriate model, the transport is diffusive in the long time limit provided that the temperature is not too close to the DNA melting (denaturation) point. Despite the fact that the diffusive charge migration can naturally be expected, its mechanism is quite peculiar. Due to the electron-phonon interaction, positive fluctuations of the vibrational subsystem act as potential barriers for the charge, so that the latter can be confined between two adjacent fluctuations during substantial part of their oscillation periods. Thus, the vibrational subsystem sets a clock for the electronic one and determines the regime of the charge migration. In other words, the dynamics of the charge density is governed by that of the vibrational subsystem: the charge transfer mechanism is fluctuation-assisted. As long as the mechanical subsystem fluctuations have finite size and time scales, the charge migration regime is diffusive in the long time limit. At shorter times (that are on the order of the characteristic timescale of the vibrational subsystem (few $ps$ at $300$ K)), the charge transport manifests traces of sub-diffusive regime. As the temperature approaches the critical one, at which the DNA undergoes a phase transition and denaturates, mechanical fluctuations become scale free. We showed that in this case the charge migration regime can also become anomalous, in particular, sub-diffusive even in the long time limit.

Summarizing, we would like to note that we considered a particular model of an organic macromolecular system in which electronic and vibrational degrees of freedom are strongly coupled due to a very specific mechanical deformations: the stretching $y_n$ of the H-bond between two DNA bases within a pair. However, the electron-phonon coupling term employed in the model Hamiltonian has a very generic form. Therefore, for another system $y_n$ can represent a very different generalized or phenomenological coordinate characterizing the dynamics of the vibrational subsystem, leaving our qualitative reasoning valid. We believe therefore that our proposed mechanism of fluctuation-assisted charge migration can be relevant for other organic systems with strong electron-phonon coupling.

\begin{acknowledgments}

This work was partially supported by CNPq, CAPES (grant PVE-A121) and FINEP (Federal Brazilian Agencies), as well as FAPEAL (Alagoas State Agency) and Spanish MINECO grants MAT2013-46308 and MAT2016-75955. AVM is grateful to the Universidade Federal de Alagoas (were a part of this work has been carried out) for hospitality. The authors would like to thank V. A. Malyshev for critical reading of the manuscript and also acknowledge the crucial contribution of J. Mun\'arriz to numerical simulations.

\end{acknowledgments}

\bibliographystyle{apsrev4-1}
\bibliography{polaron_formation}

\end{document}